\documentclass[11pt,a4paper]{article} 
\usepackage{jheppub}
\pdfoutput=1

\usepackage[dvipsnames]{xcolor}

\usepackage{tikz}
\usetikzlibrary{decorations.pathmorphing,decorations.pathreplacing,calligraphy,calc,cd,arrows.meta,patterns}
\usepackage{csquotes}
\usepackage[compat=1.1.0]{tikz-feynman}
\tikzfeynmanset{warn luatex=false}
\usepackage{contour}

\definecolor{cardinal}{rgb}{0.63,0.04,0.06}
\definecolor{ceruleanblue}{rgb}{0.16, 0.32, 0.75}
\definecolor{myblue}{rgb}{0.04,0.16,0.72}

\hypersetup{
       colorlinks=true,
       urlcolor=myblue, 
       citecolor=cardinal, 
       linkcolor=myblue 
}

\newcommand{\la}{\langle}
\newcommand{\ra}{\rangle}

\newcommand{\geps}{\gamma_E}
\newcommand{\pperp}{{\text{\tiny$\perp$}}}
\newcommand{\ma}{\mathcal{A}}
\newcommand{\mm}{\mathcal{M}}
\newcommand{\ml}{\mathsf{L}}
\newcommand{\msf}{\mathsf{M}}
\newcommand{\trfive}{\mathrm{tr}_5}
\newcommand{\geu}{\gamma_{\rm E}}
\newcommand{\zb}{\bar{z}}
\newcommand{\kallen}{K\"all\'en}
\newcommand{\veps}{\varepsilon}
\newcommand{\ham}{\hat{\rm H}}
\newcommand{\tp}{\mathsf{TP}}
\newcommand\mi[1]{\mathcal{I}_{#1}}

\usetikzlibrary{fadings}
\tikzfading[name=fade right,
right color=transparent!0,
left color=transparent!100]

\usetikzlibrary{decorations.markings}
\tikzset{
    crossdot/.style={circle, draw, inner sep=0pt, minimum size=7pt, path picture={
            \draw[black]
                (path picture bounding box.north west) -- (path picture bounding box.south east)
                (path picture bounding box.north east) -- (path picture bounding box.south west);
}}}

\tikzset{
    partial ellipse/.style args={#1:#2:#3}{
        insert path={+ (#1:#3) arc (#1:#2:#3)}
}}

\tikzset{shorten <>/.style={ shorten >=#1, shorten <=#1 }}
\tikzset{gradRtoB/.style={
    postaction={
        decorate,
        decoration={
            markings,
            mark=at position \pgfdecoratedpathlength-0.5pt with {\arrow[blue,line width=#1] {}; },
            mark=between positions 0 and \pgfdecoratedpathlength-0pt step 0.5pt with {
                \pgfmathsetmacro\myval{multiply(divide(
                    \pgfkeysvalueof{/pgf/decoration/mark info/distance from start}, \pgfdecoratedpathlength),100)};
                \pgfsetfillcolor{choral!\myval!cardinal!};
                \pgfpathcircle{\pgfpointorigin}{#1};
                \pgfusepath{fill};}
}}}}

\tikzset{
    vector/.style={
        decoration={snake, aspect=0.75, mirror, segment length=2mm},
        decorate
    },
    photon/.style={
        decorate,
        decoration={snake, amplitude=1pt, segment length=6pt}
    },
    graviton/.style={
        decorate,
        decoration={snake, amplitude=.4mm, segment length=1.5mm, pre length=.5mm, post length=.5mm},
        double
    }
}

\title{The classical two-loop amplitude at leading logarithmic accuracy for graviton emission in multi-Regge kinematics}

\preprint{\begin{minipage}[t]{8cm}\begin{flushright}
CERN-TH-2026-266 
\end{flushright}\end{minipage}}

\author[1]{Federico Buccioni,}
\author[2]{Vittorio Del Duca,}
\author[2,3]{Emanuele Rosi}
\affiliation[1]{CERN, Theoretical Physics Department, 1211 Geneva 23, Switzerland}
\affiliation[2]{INFN, Laboratori Nazionali di Frascati, 00044 Frascati (RM), Italy}
\affiliation[3]{Dipartimento di Fisica, Universit\`a di Roma ``La Sapienza'',\\ \hspace{5pt} Piazzale Aldo Moro, 2, 00185 Roma RM, Italy}
\emailAdd{federico.buccioni@cern.ch}
\emailAdd{Vittorio.DelDuca@lnf.infn.it}
\emailAdd{emanuele.rosi@uniroma1.it}
\date{\today}

\abstract{
    Using the gravitational shockwave formalism, we compute the classical part of the $\mathcal{O}(G^{7/2})$ five-point amplitude at leading logarithmic accuracy describing the emission of a graviton in the collision of two massless scalars in multi-Regge kinematics (MRK).
    The result splits into a two-particle reducible term and a three-particle reducible one, the latter being the radiative counterpart of the $H$-diagram with the emission taken in MRK limit.
    The same amplitude can be equivalently calculated as the ultrarelativistic limit of the corresponding massive amplitude by applying the optical theorem according to the conjectured universality of the leading logarithmic amplitudes in classical gravity.
}

\begin{document}

\maketitle


\section{Introduction}

The direct detection of gravitational waves~\cite{LIGOScientific:2016aoc} inaugurated a new era of precision science concerning the relativistic two-body problem.
The LIGO-Virgo-KAGRA collaboration has by now catalogued several hundred compact-binary coalescences~\cite{LIGOScientific:2026sit}, and the next generation of observatories~\cite{ET:2019dnz,LISA:2024hlh} will substantially enlarge the accessible space of parameters of the binary systems.
Matching the new precision requires analytic control of the two-body dynamics well beyond what is presently available~\cite{LISAConsortiumWaveformWorkingGroup:2023arg}.
A considerable effort has gone into weak-field expansions of the binary dynamics and into schemes that resum them~\cite{Damour:2022ybd,Long:2024ltn,Long:2025nmj}.
The scattering problem, which is related to the bound one by analytic continuation~\cite{Kalin:2019rwq,Kalin:2019inp,Cho:2021arx}, admits a post-Minkowskian (PM) expansion whenever the impact parameter is much larger than the radii of the two bodies.
Scattering observables in this expansion follow from a number of equivalent approaches, {\it e.g.} the worldline effective field theory~\cite{Kalin:2020mvi}, WQFT~\cite{Mogull:2020sak,Jakobsen:2022psy} and amplitude-based methods~\cite{Bjerrum-Bohr:2018xdl,Cheung:2018wkq,Bern:2019crd} such as the eikonal operator~\cite{DiVecchia:2021bdo,DiVecchia:2022piu,Georgoudis:2023eke}, the $\hat{\rm N}$ operator~\cite{Brandhuber:2025igz,Damgaard:2023ttc,Damgaard:2021ipf,Gonzo:2026yha,Brandhuber:2026dkd} and the KMOC formalism~\cite{Kosower:2018adc,Cristofoli:2021vyo}.
The binary dynamics is by now known through 4PM order, that is, through three-loop four-point amplitudes~\cite{Bern:2021dqo,Dlapa:2021npj,Bern:2021yeh,Dlapa:2021vgp,Dlapa:2022lmu,Jakobsen:2023ndj,Jakobsen:2023hig,Damgaard:2023ttc,Dlapa:2024cje}, and several results are already available at 5PM, specifically the full dynamics at the first order in Self-Force (SF) expansion and the conservative part at 2SF~\cite{Driesse:2024xad,Driesse:2026qiz,Bern:2026oqp,Bern:2025wyd}.
The scattering waveform is instead governed by the radiative five-point amplitude, with one graviton emitted in the final state.
Its tree-level limit, first obtained in~\cite{Kovacs:1977uw,Kovacs:1978eu}, has been rederived and extended with modern techniques~\cite{Mougiakakos:2021ckm,Jakobsen:2021smu,Jakobsen:2021lvp,Cristofoli:2021vyo,DeAngelis:2023lvf,Aoude:2023dui,Brunello:2025cot}, and both the one-loop amplitude and the waveform it generates are known in considerable detail~\cite{Brandhuber:2023hhy,Herderschee:2023fxh,Georgoudis:2023lgf,Georgoudis:2023ozp,Georgoudis:2023eke,Elkhidir:2023dco,Caron-Huot:2023vxl,Bini:2023fiz,Brandhuber:2023hhl,Bohnenblust:2023qmy,Bini:2024rsy,Bohnenblust:2025gir,Brunello:2025eso}, with partial information at higher orders~\cite{Alessio:2024onn,Bini:2024ijq,Georgoudis:2025vkk,Bini:2026vaq,Alessio:2025flu}.

Several reasons motivate the study of gravitational scattering in the high-energy limit. The PM series is organised as an expansion in $GM/b$, with $b$ the impact parameter and $M$ a mass scale, usually taken to be the total mass of the system.
There are reasons to believe that the relevant scale is instead related to the square root of the invariant mass, whose ultrarelativistic limit scales as $\gamma^{1/2}$, with $\gamma$ the relative Lorentz factor of the two bodies.
The effective expansion parameter then increases in the high-velocity limit, and the PM series must be resummed.
The same observation accounts for the non-commutativity of the $\gamma\to\infty$ and $G\to0$ expansions, since both variables enter through a single dimensionless combination.
Indeed, the PM prediction for the scattering angle converges slowly towards the NR benchmarks, and so do the resummation models built from it, see {\it e.g.}~\cite{Swain:2024ngs}.
From 4PM onwards the scattering angle develops itself divergences in the high-energy limit~\cite{Dlapa:2022lmu,Bini:2022enm}, hence a systematic treatment of the ultrarelativistic regime is therefore needed, together with resummation strategies able to keep these divergences under control.

Long ago, Amati, Ciafaloni and Veneziano used the eikonal exponentiation to organise the classical contributions to Planckian-energy collisions of superstrings, and found in the field-theory and classical limit the first correction to massless scattering, at order $\mathcal{O}(G^3)$, together with the associated bremsstrahlung~\cite{Amati:1990xe,Amati:1993tb,Amati:2007ak,Ciafaloni:2015xsr}.
That correction is encapsulated in the $H$-diagram, which has since been studied in great detail, proving that the scattering deflection arising from it is universal, hence unaffected by the massive or massless nature of the colliding bodies~\cite{Damour:2017zjx,DiVecchia:2020ymx,Bern:2020gjj}.
Later, a forward soft-collinear effective theory for massless collisions in gravity~\cite{Rothstein:2024nlq} revealed that the $H$-diagram is only the first element of a tower of multi-$H$ diagrams, which control the $(2n+1)$PM amplitudes, and the double-$H$ diagram at 5PM has been computed explicitly~\cite{Alessio:2025isu,Alessio:2026bdi}.
It can be proved via KMOC formalism~\cite{Kosower:2018adc} that none of the diagrams in this sequence contributes to classical observables directly, therefore proving the cancellation of the leading logarithmic parts of the amplitudes from the observables.
On the other hand, crossing symmetry ties them to the conservative amplitude, from which non-trivial information can be extracted, see {\it e.g.}~\cite{Amati:1990xe,Alessio:2025isu}, while a complete PM resummation of the tower would give a great deal more.
Recently, 1SF amplitudes have been resummed in the high-energy limit within a SCET framework in~\cite{Saavedra:2026ttk}, where the author furnished a recipe to calculate the 1SF scattering deflection to all orders in Newton's constant.

Computations in the Regge limit are highly simplified by the natural factorisation of the amplitude into a longitudinal dynamics and a purely transverse one in $D-2$ dimensions.
This structure is familiar from QCD, where the BFKL equation~\cite{Kuraev:1976ge,Kuraev:1977fs,Lipatov:1976zz,Balitsky:1978ic} and reggeon effective field theory~\cite{Lipatov:1989bs,DelDuca:1995hf,Gribov:2003nw} resum logarithms of the centre-of-mass energy, with remarkable integrable structures emerging in the process~\cite{Lipatov:1993qn,Lipatov:1993yb,Lipatov:2009nt}, and where high-energy factorisation has been pushed to high loop orders and to higher multiplicity~\cite{Caron-Huot:2017fxr,Falcioni:2021buo,DelDuca:2019tur,Caron-Huot:2020vlo,DelDuca:2021vjq,Caola:2021izf,Falcioni:2021dgr,Buccioni:2024gzo,Abreu:2024xoh}.
Gravity inherits a good deal of this machinery, starting from the graviton Regge trajectory and the multi-Regge emission vertex, the latter being originally derived by Lipatov~\cite{Lipatov:1982it} and having recently found a variety of applications and extensions~\cite{Raj:2024xsi,Raj:2025hse,Barcaro:2025ifi,Raj:2023irr}.
Eikonal and Wilson-line descriptions of gravitational scattering at high energy~\cite{Naculich:2011ry,White:2011yy,Melville:2013qca}, together with effective-theory treatments of soft and Glauber gravitons~\cite{Beneke:2012xa,Chiu:2012ir,Okui:2017all}, complete the picture.
In particular, the shockwave formulation developed in QCD~\cite{Mueller:1994jq,Balitsky:1995ub,Jalilian-Marian:1996mkd,Jalilian-Marian:1997jhx,Kovchegov:1999yj,Iancu:2001ad,Caron-Huot:2013fea}, in which the resummation of rapidity logarithms becomes an evolution equation for correlators of Wilson lines, has recently been transposed to gravity~\cite{Alessio:2026bdi,Bohnenblust:2026ujk}.

In this paper we use that formulation to compute the classical part of the $\mathcal{O}(G^{7/2})$ five-point amplitude at leading logarithmic accuracy for graviton bremsstrahlung in multi-Regge kinematics (MRK).
The two colliding particles are described by Wilson lines, expanded in reggeon fields and evolved in rapidity by a boost Hamiltonian, while the graviton emission is encoded through the operator product expansion between the emission operator and the reggeon fields, whose leading term is the Lipatov's gravitational central emission vertex (CEV).
The result splits into two structures of distinct physical origin: a two-particle reducible term, which is the convolution in the transverse plane of a quantum correction with a classical exchange and therefore does not contribute to classical scattering data, and a three-particle reducible one, which is the radiative counterpart of the $H$-diagram and carries the classical radiative information.

The paper is organised as follows.
In Section~\ref{sec:five-point-kinematics} we set up the five-point kinematics and the multi-Regge parametrisation, and we fix the structure of the perturbative expansion in the MRK parameter by empirical ansatz.
In Sec.~\ref{sec:shockwave-formalism} we develop the ingredients of shockwave evolution required for classical graviton emission.
Section~\ref{sec:five-point-amplitude} contains our main result and its derivation, namely the classical two-loop five-point amplitude at leading logarithmic accuracy, together with its soft limit consistency check.
We conclude in Section~\ref{sec:conclusions}.

\paragraph{Conventions:} We describe flat space using the mostly minus metric $\eta={\rm diag}(+,-,-,-)$, and the convention for the gravitational coupling constant is fixed by $\kappa^2=8\pi G$.
We work in the dimensional regularisation scheme with $D=4-2\veps$ dimensions and renormalisation scale $\mu$.
The light-cone coordinates are defined by $x^{\mu} = (x^+,x^-,\vec{x})$, with $x^\pm = x^0\pm x^3$ and $\vec{x}$ belonging to the Euclidean space of dimensionality $d=D-2$ that we label as transverse (sub)space.
We also use $x_\pperp = (0,0,\vec{x})$, so that $x_\pperp^2=-\vec{x}^{\,2}$, and latin indices implicitly project onto the transverse subspace $x^i = x_\pperp^i$.
The loop integration measure is then defined as
\begin{equation}
       \int_\ell = \frac{1}{2}\int \hat{d}\ell^{+} \int\hat{d}\ell^{-} \int_{\ell_\pperp}\,, \hspace{20pt} \int_{\ell_\pperp} = (\mu^2 e^{\geu})^\veps \int \hat{d}^d\ell_\pperp
\end{equation}
with $\hat{d}\ell^\mu = {d\ell^\mu}/{2\pi}$.
Likewise, we normalise the delta functions $\hat{\delta}^{(N)}(f) = (2\pi)^N {\delta}^{(N)}(f)$.
When specialising to $d=2$, we express the transverse momenta through one complex coordinate $\zeta_p = p^1 + ip^2$.


\section{Five-point kinematics and the multi-Regge limit}
\label{sec:five-point-kinematics}
The object of this work is the five-point amplitude describing the emission of a graviton in the high-energy collision of two massless particles.
We consider the Einstein-Hilbert lagrangian coupled to two massless scalar fields, which describe the ultrarelativistic particles.
We work in the classical limit, in which the two massless particles are considered classical sources, while the gravitons are soft and their momentum is taken proportional to $\hbar$.
Before deriving the amplitude from the shockwave formalism, we introduce the kinematics of a five-point process, specialising to the physical production of a graviton in the multi-Regge region.

\subsection{The massless five-point kinematics}
\label{sec:five-point-process}
\begin{figure}[t!]
       \centering
       \begin{tikzpicture}[baseline=(blob.center),scale=1,transform shape]
              \begin{feynman}
                     \vertex[dot] (blob) {};
                     \vertex[right=4cm of blob] (outk) {$P_4^\mu = q_1^\mu + q_2^\mu$};
                     \vertex[above=35pt of blob] (Aup) {};
                     \vertex[below=35pt of blob] (Alow) {};
                     
                     \vertex[right=4cm of Aup] (outUp) {$P_5^\mu = p_1^\mu - \frac{q_1^\mu}{2}$};
                     \vertex[right=4cm of Alow] (outLow) {$P_3^\mu = p_2^\mu - \frac{q_2^\mu}{2}$};
                     \vertex[left=4cm of Aup] (inUp) {$P_1^\mu = - p_1^\mu - \frac{q_1^\mu}{2}$};
                     \vertex[left=4cm of Alow] (inLow) {$P_2^\mu = - p_2^\mu - \frac{q_2^\mu}{2}$};
        
                     \diagram{
                            {
                                   (blob) -- [edges=graviton] (outk),
                                   (inUp) -- [RoyalBlue!100, ultra thick] (Aup) -- [RoyalBlue!100, ultra thick] (outUp),
                                   (inLow) -- [cardinal!80, ultra thick] (Alow) -- [cardinal!80, ultra thick] (outLow)
                            }
                     };
                     \vertex[dot, style={fill=gray!15, thick, shape=ellipse, minimum width=60pt, minimum height=80pt}, left=4cm of outk] (blob2) {\large$\mm_5^\lambda$};
              \end{feynman}
       \end{tikzpicture}
       \caption{Kinematics of the massless five-point amplitude with all outgoing convention. The average momenta of the massless and spinless particles are named $p_i$, while $q_i$ their difference. We use the double wavy line to describe a graviton in the multi-Regge region.}
       \label{fig:kinematics}
\end{figure}
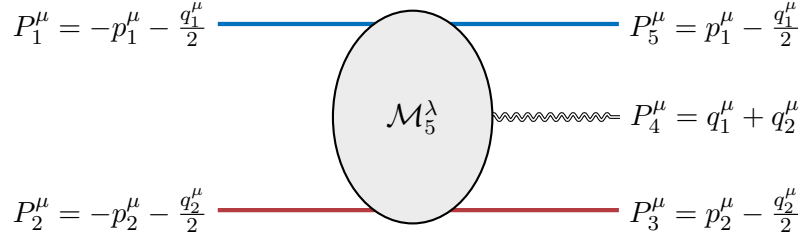

We denote by $\mm_{2\to3}^{\lambda} = \mm_{2\to3}(P_1,P_2,P_3,P_4^\lambda,P_5)$ the amplitude for the scattering of two massless scalars with the emission of bremsstrahlung radiation, where $\lambda = \pm$ labels the helicity of the outgoing graviton.
As depicted in Figure~\ref{fig:kinematics}, we adopt the all-outgoing convention,
\begin{equation}
\label{eq:five-point-process}
P_1^\mu + P_2^\mu + P_3^\mu + P_4^\mu + P_5^\mu = 0\,, \qquad P_i^2 = 0\,.
\end{equation}
A convenient set of independent parity-even invariants is
\begin{equation}
\label{eq:five-point-invariants}
s_{12}, s_{23}, s_{34}, s_{45}, s_{51},
\qquad s_{ij}\equiv(P_i+P_j)^2=2P_i\cdot P_j\,,
\end{equation}
and the physical $2\to3$ region considered throughout the paper is determined by
\begin{equation}
       s_{12},s_{34},s_{45} > 0 \,, \qquad \text{and} \qquad s_{23},s_{51} <0,
\end{equation}
namely the region in which $P_{1,2}$ are the initial states and $P_{3,4,5}$ the final ones.
We also define the parity-odd quantity
\begin{equation}
\label{eq:tr5-definition}
\trfive
=4i\,\epsilon_{\mu\nu\rho\sigma}
P_1^{\mu}P_2^{\nu}P_3^{\rho}P_4^{\sigma}\,,
\end{equation}
which obeys $\trfive^2=\Delta$, where $\Delta$ is the Gram determinant of the process.
This allows us to distinguish the physical region which, as we show below, features $\Delta<0$ and $\trfive$ purely imaginary.
Therefore, specifying the sign of $\trfive$ selects one of the two parity-related points associated with the same Mandelstam invariants~\cite{Gehrmann:2018yef,Chicherin:2020oor}.
To expose the classical scaling, we parametrise the external momenta as
\begin{equation}
\label{eq:average-transfer-parametrisation}
\begin{aligned}
P_1&=-p_1-\frac{q_1}{2}\,,
&\qquad P_5&=p_1-\frac{q_1}{2}\,,
&\qquad P_4&\equiv k=q_1+q_2\,,\\
P_2&=-p_2-\frac{q_2}{2}\,,
&P_3&=p_2-\frac{q_2}{2}\,.
\end{aligned}
\end{equation}
Here $p_1$ and $p_2$ are the hard average momenta of the two scattered particles, whereas $q_1$ and $q_2$ are momentum transfers. The on-shell conditions imply
\begin{equation}
\label{eq:on-shell-average-transfer}
p_i\cdot q_i=0\,,\qquad
p_i^2=-\frac{q_i^2}{4}\,,\qquad
2q_1\cdot q_2=-q_1^2-q_2^2\,,
\end{equation}
so that the independent invariants become
\begin{equation}
\label{eq:invariants-average-transfer}
\begin{aligned}
s_{12}&=2p_1\cdot p_2+p_1\cdot q_2+p_2\cdot q_1
       -\frac{q_1^2+q_2^2}{4}\,,
\qquad s_{51}=q_1^2\,,
\qquad s_{23}=q_2^2\,,\\
s_{45}&=2p_1\cdot q_2-\frac{q_1^2-q_2^2}{2}\,,
\qquad s_{34}=2p_2\cdot q_1-\frac{q_2^2-q_1^2}{2}\,,
\qquad \trfive=4i\epsilon_{\mu\nu\rho\sigma}p_1^{\mu}p_2^{\nu}q_1^{\rho}q_2^{\sigma}\,.
\end{aligned}
\end{equation}
Restoring powers of $\hbar$, the transfer momenta scale as $q_i=\mathcal{O}(\hbar)$ while the hard momenta are of order $p_i=\mathcal{O}(\hbar^0)$.
This parametrisation thus makes manifest the hierarchy between classical, superclassical and quantum terms, which will be used in Sec.~\ref{sec:five-point-amplitude}.

\subsection{Multi-Regge limit}\label{sec:multi-regge-limit}

Multi-Regge kinematics isolates the high-energy regime in which emissions are strongly ordered in rapidity, while their transverse momenta remain comparable.
For the process~\eqref{eq:five-point-process}, it corresponds to the ordering
\begin{equation}
\label{eq:mrk-lightcone-ordering}
P_5^+\gg P_4^+\gg P_3^+\,,
\qquad
P_5^-\ll P_4^-\ll P_3^-\,,
\qquad
|P_{3\perp}|\sim |P_{4\perp}|\sim |P_{5\perp}|\,,
\end{equation}
or, equivalently, in terms of the final-state rapidities,
\begin{equation}
\label{eq:mrk-rapidity-ordering}
y_5\gg y_4\gg y_3\,,
\qquad y_i\equiv\frac{1}{2}\log\frac{P_i^+}{P_i^-}\,.
\end{equation}
Following the parametrisation of refs.~\cite{Caron-Huot:2020vlo,Chicherin:2020oor}, we introduce a positive parameter $x$ through
\begin{equation}
\label{eq:mrk-scaling}
s_{12}=\frac{s}{x^2}\,,\qquad
s_{45}=\frac{s_1}{x}\,,\qquad
s_{34}=\frac{s_2}{x}\,,\qquad
s_{51}=t_1\,,\qquad
s_{23}=t_2\,,
\end{equation}
such that the MRK limit is $x\to0^+$ while the reduced variables $s,s_1,s_2,t_1,t_2$ are held fixed.
In terms of the invariant products, we see from~\eqref{eq:invariants-average-transfer} that the MRK limit corresponds to $p_1\cdot p_2 \gg p_1\cdot q_2 \sim p_2\cdot q_1 \gg q_i\cdot q_j$, which coincides with the classical limit, since $q_i = \mathcal{O}(\hbar)$ and $p_i = \mathcal{O}(\hbar^0)$.
The parity-odd invariant also simplifies in this limit. At leading power,
\begin{equation}
\label{eq:tr5-mrk}
\trfive^2
=\frac{s_1^2s_2^2}{x^4}
\lambda\!\left(1,-\frac{s t_1}{s_1s_2},-\frac{s t_2}{s_1s_2}\right)
+\mathcal{O}(x^{-3})\,,
\end{equation}
where $\lambda(a,b,c)=a^2+b^2+c^2-2ab-2ac-2bc$ is the \kallen{} function.
Equation~\eqref{eq:tr5-mrk} exhibits the high-energy factorisation explicitly, already well known in Yang-Mills theories~\cite{Kuraev:1976ge,Kuraev:1977fs}.
The $x^{-4}$ high-energy enhancement is factored out of the \kallen{} function, which now depends solely on the transverse momenta.
Indeed, we prove below that the entries of the \kallen{} function in eq.~\eqref{eq:tr5-mrk} can be rewritten uniquely in terms of $q_{1\pperp}$ and $q_{2\pperp}$.

Following~\cite{Chavez:2012kn}, it is convenient to parametrise the transverse dynamics through a single scale $Q$,
\begin{equation}
\label{eq:transverse-scale}
Q^2\equiv -k_\pperp^2 = -(q_1+q_2)_\pperp^2
=\frac{s_1s_2}{s}+\mathcal{O}(x)\,,
\end{equation}
which will be factored out of the Feynman integrals, together with two dimensionless scales $u,v$ related respectively to $q_{1\pperp}$ and $q_{2\pperp}$,
\begin{equation}
\label{eq:uv-variables}
u\equiv - \frac{q_{1\pperp}^2}{Q^2}=z\zb\,,
\qquad
v\equiv - \frac{q_{2\pperp}^2}{Q^2}=(1-z)(1-\zb)\,,
\end{equation}
that we express in terms of two complex numbers $z,\zb$.
Since at leading power $t_i\simeq q_{i\pperp}^2$, we get
\begin{equation}
\label{eq:transfers-complex-variables}
t_1\simeq-\frac{s_1s_2}{s}z\zb\,,
\qquad
t_2\simeq-\frac{s_1s_2}{s}(1-z)(1-\zb)\,.
\end{equation}
Solving the system~\eqref{eq:uv-variables} for $z$ and $\zb$ yields
\begin{equation}
\label{eq:zbarz-from-uv}
z=\frac{1}{2}\left(1+u-v+\sqrt{\lambda(1,u,v)}\right)\,,
\qquad
\zb=\frac{1}{2}\left(1+u-v-\sqrt{\lambda(1,u,v)}\right)\,.
\end{equation}
In the physical region, the Cauchy--Schwarz inequality gives
\begin{equation}
\label{eq:kallen-transverse}
\lambda(1,u,v)
=\frac{4}{Q^4}\Big[ (q_{1\pperp}\cdot k_\pperp)^2 - q_{1\pperp}^2 k_\pperp^2 \Big]
=(z-\zb)^2\leq0\,,
\end{equation}
so that $z$ and $\zb$ are complex conjugates, except in the collinear limit, where the \kallen{} function vanishes.
We fix their labelling, and hence the sign of the parity-odd invariant, by choosing
\begin{equation}
\label{eq:tr5-zbarz}
\trfive = \frac{s_1 s_2}{x^2}(z-\zb)+\mathcal{O}(x^{-1})\,.
\end{equation}
The pair $(z,\zb)$ encodes the whole non-trivial transverse dependence of the kinematics in the high-energy and MRK limit.
As a consequence of the choice~\eqref{eq:transfers-complex-variables} and of the sign of $\trfive$ in~\eqref{eq:tr5-zbarz}, we get the following identities\footnote{In deriving these identities we are free to fix a direction in the transverse plane. Following App.~A of~\cite{Buccioni:2024gzo}, we take $k_\pperp$ on the negative real axis, {\it i.e.} $k_\pperp^{i} = -(Q,0)$.}
\begin{equation}\label{eq:complez-zzb-q1q2k}
       \zeta_{k} = -\sqrt{\frac{s_1 s_2}{s}} \,,
       \qquad 
       \zeta_{q_1} = -\sqrt{\frac{s_1 s_2}{s}}\,\zb \,,
       \qquad
       \zeta_{q_2} = -\sqrt{\frac{s_1 s_2}{s}}(1-\zb) \,.
\end{equation}
Finally, we introduce the graviton polarisation tensors in the light-cone gauge~\cite{Alessio:2026bdi} as the double copy of the Yang-Mills polarisation vectors~\cite{Bern:2019crd,DelDuca:1995zy},
\begin{equation}\label{eq:grap-pols}
       \epsilon_{\mu\nu}^{(\lambda)}(k) = \epsilon_{\mu}^{(\lambda)}(k)\epsilon_{\nu}^{(\lambda)}(k) \,,
\end{equation}
where
\begin{equation}\label{eq:ym-pols}
       \epsilon_\oplus^{\mu}(k) = \frac{1}{\sqrt{2}}\left(\frac{2\zeta_k}{k^{-}},0,1,i\right),
\end{equation}
and $\epsilon_\mu^\ominus = \epsilon_{\mu}^{\oplus\,*}$.

\subsection{Five-point amplitude structure in MRK}
Given the expansion of the kinematic invariants in the small MRK parameter $x$, we can write a formal series for the perturbative $2\to3$ amplitude.
We first expand in the number of loops,
\begin{equation}
       \mm_{2\to3}^\lambda = \sum_{l=0}^{\infty} \mm_{2\to3}^{(l),\lambda}\,,
\end{equation}
with $\mm_{2\to3}^{(l),\lambda}\propto\kappa^{3+2l}$.
Then we take the limit $x\to0^+$, for which we ansatz the following asymptotic expansion
\begin{equation}\label{eq:m5ll}
       \mm_{2\to3}^{(l),\lambda} \sim \sum_{n=0}^{\infty} \mm_{2\to3}^{(l,n),\lambda} = \frac{\kappa^{3+2l}}{x^{4+2l}} \sum_{n=0}^{\infty} \left[\msf^{(l,n),\lambda}_{2\to3} \cdot x^n (\log x)^{\lfloor n/2 \rfloor} + \;\cdots\; \right] ,
\end{equation}
which we justify at the end of Section~\ref{sec:shockwave-formalism}.
The dots in the square brackets denote terms with the same power in $x$, but logarithmically suppressed, namely $\sim x^n\log^{\lfloor n/2 \rfloor-k}(x)$ with $k\ge1$.
In the next section we describe the shockwave formalism at leading logarithmic accuracy, which is what we need to build the tower of coefficients $\msf_{2\to3}^{(l,n),\lambda}$ with even $n$.


\section{Review of the gravitational shockwave formalism}\label{sec:shockwave-formalism}
This section gives a brief overview of the shockwave formalism, which we use to derive the $2\to3$ amplitudes in MRK and has been formulated in~\cite{Alessio:2026bdi} and used to construct conservative amplitudes and the tree-level MRK radiative amplitude.
The shockwave formalism for gravity builds on the widely used shockwave formalism for QCD~\cite{Balitsky:1995ub,Mueller:1994jq,Kovchegov:1999yj,Jalilian-Marian:1996mkd,Jalilian-Marian:1997jhx,Iancu:2001ad,Caron-Huot:2013fea} and describes the forward collision of two massless particles, with the associated $2\to2$ amplitude related to the time-ordered correlator between two Wilson lines by~\cite{Naculich:2011ry,Melville:2013qca,White:2011yy}
\begin{equation}\label{eq:amplitude-correlator-2in2-divergent}
    i{\rm FT}\big[\mm_{2\to2}\big] (s,b) = \Big\la \tilde\Phi_2(b_{2}) \, \tilde\Phi_1(b_{1}) \Big\ra\,,
\end{equation}
each one defined as
\begin{equation}\label{eq:wl-definition}
    \tilde\Phi_a(b_a) = \exp\left\{ \frac{i\kappa}{\sqrt{2}}\int_{\mathbb{R}} d\tau\,P_a^\mu\,P_a^\nu\,h_{\mu\nu}\left(x_a^\mu(\tau)\right) \right\} \,.
\end{equation}
We use $\la\hspace{3pt}\cdot\hspace{3pt}\ra$ as a shortcut for the time-ordered vacuum expectation value, namely $\la 0 |\mathbf{T} \hspace{3pt}\cdot\hspace{3pt} | 0 \ra$.
Here $x_a^\mu(\tau) = -P_a^\mu \tau + b_{a\pperp}$ is the straight-line trajectory of the body labelled by $a=1,2$, $b_{a\pperp}$ is its impact parameter and $P_a^\mu$ are the momenta introduced in equation~\eqref{eq:average-transfer-parametrisation}, that we choose to be $P_1^\mu=-\sqrt{s_{12}}\,\big(1,0,\vec{0}\big)$ and $P_2^\mu=-\sqrt{s_{12}}\,\big(0,1,\vec{0}\big)$.
The presence of a source pointwise-localised on $x_a^\mu(\tau)$ reflects the eikonal approximation underlying the formalism.
The correlator~\eqref{eq:amplitude-correlator-2in2-divergent} diverges in the rapidity difference between the two bodies, which is itself a formally infinite quantity.
This divergence is not renormalised in the usual dimensional regularisation scheme, given its geometrical nature.
Thus one has to introduce a new parameter $\eta$ describing the rapidity of the Wilson line $\tilde\Phi_1$, which is now named $\tilde{\Phi}^\eta_1$.
One of the two lines must be rapidity-renormalised, evolving its value up to the rapidity of the other line, via the evolution equation
\begin{equation}\label{eq:rrge}
    \frac{d\tilde\Phi_a^\eta}{d\eta} + \ham\,\tilde\Phi_a^\eta = 0 \,,
\end{equation}
and the {\it boost Hamiltonian} $\ham$ that we introduce below.\footnote{In the same spirit, the forward-scattering effective theory of gravity introduced in~\cite{Rothstein:2024nlq} renormalises the divergent amplitude through a Rapidity Renormalisation Group Equation (RRGE), which acts either on the Wilson lines or on the soft operator that mediates the interactions among the lines~\cite{Chiu:2012ir}.}
Therefore, the renormalised amplitude is computed using
\begin{equation}\label{eq:amplitude-correlator-2in2}
    i{\rm FT}\big[\mm_{2\to2}\big] (s,b) = \Big\la \tilde\Phi_2(b_{2}) \, e^{-\ham\ml} \, \tilde\Phi_1(b_{1}) \Big\ra = \Big\la \tilde\Phi_2(b_{2}) \, \tilde\Phi_1^\ml(b_{1}) \Big\ra\,,
\end{equation}
where now $\ml \simeq \log(-s/t)$ is the rapidity gap and $\exp(-\ham\ml)$ is called evolution operator.
In what follows, we review the key elements needed to compute the amplitudes, referring the reader to~\cite{Alessio:2026bdi} for technicalities and a more detailed discussion.

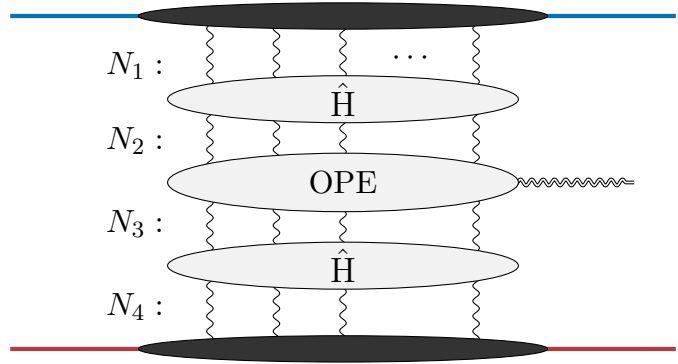
\begin{figure}[t!]
       \vspace{10pt}
       \centering
       \begin{tikzpicture}[baseline=(init.base),scale=1.1,transform shape]
              \begin{feynman}
                     \vertex[dot] (init) {};
                     \vertex[right=4cm of init] (in);
                     \vertex[left=4cm of init] (out);

                     \vertex[below=4cm of init] (kmid);
                     \vertex[right=4cm of kmid] (in2);
                     \vertex[left=4cm of kmid] (out2);

                     \vertex[left=1.6cm of kmid] (k0);
                     \vertex[left=0.8cm of kmid] (k1);
                     \vertex[right=1.6cm of kmid] (k2);

                     \vertex[left=1.6cm of init] (i0);
                     \vertex[left=0.8cm of init] (i1);
                     \vertex[right=1.6cm of init] (i2);

                     \vertex[below=2cm of init] (center);
                     \vertex[right=3.5cm of center] (kout);

                     \vertex[below=.5cm of init] (sta);
                     \vertex[right=0.45 of sta] (dots) {$\cdots$};
                     \vertex[right=1.5cm of out] (outN);
                     \vertex[below=0.3cm of outN] (n1) {$N_1:$};
                     \vertex[below=1.2cm of outN] (n2) {$N_2:$};
                     \vertex[below=2.2cm of outN] (n3) {$N_3:$};
                     \vertex[below=3.2cm of outN] (n4) {$N_4:$};
              
                     \diagram{
                            {
                                   (in) -- [RoyalBlue!100, ultra thick] (init) -- [RoyalBlue!100, ultra thick] (out),
                                   (in2) -- [cardinal!80, ultra thick] (kmid) -- [cardinal!80, ultra thick] (out2),
                                   (init) -- [edges=photon] (kmid),
                                   (i0) -- [edges=photon] (k0),
                                   (i1) -- [edges=photon] (k1),
                                   (i2) -- [edges=photon] (k2),
                                   (center) -- [edges=graviton] (kout)
                            },
                     };
                     \vertex[dot, style={fill=black!80, shape=ellipse, minimum width=140pt, minimum height=9pt}, right=0cm of init] {};
                     \vertex[dot, style={fill=black!80, shape=ellipse, minimum width=140pt, minimum height=9pt}, below=4cm of blob] {};
                     \vertex[dot, style={fill=gray!10, shape=ellipse, minimum width=120pt, minimum height=15pt}, below=1cm of init] {\large$\ham$};
                     \vertex[dot, style={fill=gray!10, shape=ellipse, minimum width=120pt, minimum height=20pt}, below=2cm of init] (ope) {OPE};
                     \vertex[dot, style={fill=gray!10, shape=ellipse, minimum width=120pt, minimum height=15pt}, below=3cm of init] {\large$\ham$};
              \end{feynman}
       \end{tikzpicture}
       \caption{The five-point amplitude $\mm_{2\to3}^\lambda$ obtained from the shockwave formalism~\eqref{eq:amplitude-correlator-2in3}.
       At fixed order in perturbation theory, the blue Wilson line interacts with the gravitational field through $N_1$ reggeon fields, which evolve via $\ham \supset \hat{\mathcal{R}}_{N_1,N_2}$ into $N_2$ reggeons;
       afterwards, one selects the emission $\hat{a}^\lambda\cdot W^{N_2} \to W^{N_3}$ via the operator product expansion;
       the produced reggeons interact again via the Hamiltonian $\hat{\mathcal{R}}_{N_3,N_4}$ to regularise the rapidity divergences, and finally the remaining $N_4$ reggeons are absorbed by the red Wilson line.}
       \label{fig:5pts-from-shocks}
\end{figure}

\paragraph{Reggeon basis:}
The Wilson lines introduced in equation~\eqref{eq:wl-definition} represent coherent states of gravitons sourced by the trajectories $x_a^\mu(\tau)$.
These gravitons are referred to as {\it reggeon fields} $\tilde{W}$,
\begin{equation}\label{eq:W-field}
\tilde\Phi_1(b_1) = \hspace{3pt}
\begin{tikzpicture}[baseline=(floor.base),scale=1,transform shape]
       \begin{feynman}
              \vertex[dot] (init) {};
              \vertex[right=2.5cm of init] (in);
              \vertex[left=2.5cm of init] (out);

              \vertex[below=.5cm of init] (floor);
              \vertex[below=1.1cm of init] (kmid);
              
              \vertex[left=1.4cm of kmid] (k0);
              \vertex[left=.7cm of kmid] (k1);
              \vertex[right=1.4cm of kmid] (k2);

              \vertex[left=0.9cm of init] (i0);
              \vertex[left=0.45cm of init] (i1);
              \vertex[right=0.9cm of init] (i2);

              \vertex[right=.6cm of kmid] (etc) {};
              \vertex[above=.1cm of etc] (etc2) {$\cdots$};
 
              \diagram{
                     {
                            (in) -- [RoyalBlue!100, ultra thick] (init) -- [RoyalBlue!100, ultra thick] (out),
                            (init) -- [edges=photon] (kmid),
                            (i0) -- [edges=photon] (k0),
                            (i1) -- [edges=photon] (k1),
                            (i2) -- [edges=photon] (k2),
                     },
              };
              \vertex[dot, style={fill=black!80, shape=ellipse, minimum width=80pt, minimum height=9pt}, right=0cm of blob] (blob) {};
       \end{feynman}
\end{tikzpicture}
\hspace{3pt} = \exp\left\{ i \kappa\,\sqrt{\frac{s_{12}}{2}}\,\tilde{W}(b_1)\right\} \,,
\end{equation}
and describe the instantaneous interactions taking place on the wavefront of the shockwave produced by the massless particle.
For this reason, they depend only on the transverse dynamics.
We distinguish reggeons from the graviton emitted in MRK by using a single wavy line for the former, and a double wavy line for the latter, see {\it e.g.} Figure~\ref{fig:5pts-from-shocks}.
Moreover we define the state with $n$ reggeons to be the $n-$th order term of the Taylor expansion of the exponential~\eqref{eq:W-field}, expressed in momentum domain via Fourier transform
\begin{equation}\label{eq:wline-expanded}
       \Phi^{(n)}_a(q_a) = \frac{i^n \kappa^n}{x^n n!} \bigg(\frac{s}{2}\bigg)^{n/2} {\rm FT} \Big[ \tilde{W}^n (b_a) \Big] (q_a) \,, \hspace{25pt} n\ge1 \,.
\end{equation}
The perturbative computation of the amplitude requires knowledge of the correlator between two reggeons, which is written in momentum space as~\cite{Caron-Huot:2013fea}
\begin{equation}
    \Big\la W(r) W(r^\prime) \Big\ra = \frac{i}{r_{\pperp}^2} \hat{\delta}^{(d)}(r_{\pperp} + r_{\pperp}^\prime)  + \mathcal{O}(\kappa^2) \,.
\end{equation}
All the other correlators between an even number of reggeons are computed by applying Wick contractions.

\paragraph{Boost Hamiltonian:}
Expanding the correlator~\eqref{eq:amplitude-correlator-2in2} perturbatively requires knowledge of the interactions between the reggeons sourced by the two eikonal lines.
This information comes entirely from the bulk action and is embedded in the evolution Hamiltonian $\ham$, which is expressed perturbatively as~\cite{Alessio:2026bdi}
\begin{equation}
       \label{eq:boost-H}
       \ham = \hat{\mathcal{R}}_{1,1}^{(1)} + \hat{\mathcal{R}}_{2,2}^{(1)} + \mathcal{O}(\kappa^4) = \hspace{4pt}
       \begin{tikzpicture}[baseline=(base.center),scale=1,transform shape]
              \begin{feynman}
                     \vertex[dot] (mid) {};
                     \vertex[above=.9cm of mid] (init) {};
                     \vertex[below=.9cm of mid] (fin) {};
                     \vertex[below=.1cm of mid] (base);
              
                     \diagram{
                            (init) -- [edges=photon] (mid) -- [edges=photon] (fin)
                     };
                     \vertex[dot, style={fill=gray!10, shape=ellipse, minimum width=15pt, minimum height=15pt}, below=.9cm of init] (blob) {\small$\hat{\mathcal{R}}_{1,1}^{(1)}$};
              \end{feynman}
       \end{tikzpicture}
       \hspace{4pt} + \hspace{4pt}
       \begin{tikzpicture}[baseline=(base.center),scale=1,transform shape]
              \begin{feynman}
                     \vertex[dot] (mid2) {};
                     \vertex[right=.3cm of mid] (mid3) {};
                     \vertex[left=.3cm of mid2] (mid) {};
                     \vertex[above=.9cm of mid] (init) {};
                     \vertex[below=.9cm of mid] (fin) {};
                     \vertex[above=.9cm of mid3] (init3) {};
                     \vertex[above=.9cm of mid2] (init2) {};
                     \vertex[below=.9cm of mid3] (fin3) {};
                     \vertex[below=.1cm of mid] (base);
              
                     \diagram{
                            (init) -- [edges=photon] (mid) -- [edges=photon] (fin),
                            (init3) -- [edges=photon] (mid3) -- [edges=photon] (fin3)
                     };
                     \vertex[dot, style={fill=gray!10, shape=ellipse, minimum width=20pt, minimum height=15pt}, below=.9cm of init2] (blob) {\small$\hat{\mathcal{R}}_{2,2}^{(1)}$};
              \end{feynman}
       \end{tikzpicture}
       \hspace{5pt} + \mathcal{O}(\kappa^4) \,,
\end{equation}
where the operators $\hat{\mathcal{R}}_{i,j}^{(m)}\propto\kappa^{2m}$ act on a state composed of $j$ reggeons and return $i$ reggeons.
For example, in the diagrams of equation~\eqref{eq:boost-H}, the two operators evolve, respectively, the one- and two-reggeon states from above to below.
The operators $\hat{\mathcal{R}}_{1,1}^{(m)}$ are defined precisely as the insertion of a graviton Regge trajectory at $m$ loops.
For instance, at $m=1$ we obtain
\begin{equation}\label{eq:hatR1}
       \hat{\mathcal{R}}_{1,1}^{(1)} = - \int_{\ell_\pperp} \;\alpha^{(1)}(\ell)\,W(\ell)\,\frac{\delta}{\delta W(\ell)} \,,
\end{equation}
with $\alpha^{(1)}$ the known one-loop Regge trajectory formula of Lipatov~\cite{Lipatov:1982it}, which can be compactly rewritten as
\begin{equation}
       \alpha^{(1)}(\ell)=-\frac{\kappa^2}{4\pi}(d+1)(\ell_{\pperp}^2)^2 \int_{r_\pperp} \left[-r_\pperp^2\right]^{-1} \left[-(r_\pperp-\ell_{\pperp})^2\right]^{-1} \,.
\end{equation}
The operator $\hat{\mathcal{R}}_{2,2}^{(1)}$ is more involved, and reads
\begin{equation}\label{eq:hatr2}
    \hat{\mathcal{R}}_{2,2}^{(1)} = -\frac{\kappa^2}{8\pi} \int_{\ell_\pperp,\ell_{1\pperp},\ell_{2\pperp}} \frac{\mathcal{H}_{\mathrm{GR}}(\ell_1,\ell+\ell_1;\ell_1+\ell_2)}{\ell_{1\pperp}^2 \ell_{2\pperp}^2} W(\ell_1\!+\!\ell) W(\ell_2\!-\!\ell) \frac{\delta}{\delta W(\ell_1)}\,\frac{\delta}{\delta W(\ell_2)} \,,
\end{equation}
where $\mathcal{H}_{\mathrm{GR}}$ is the kernel of the gravitational BFKL equation, obtained by squaring the Lipatov current (CEV)~\cite{Lipatov:1982it,Raj:2025hse,Rothstein:2024nlq}.
The convention for $\mathcal{H}_{\mathrm{GR}}$ adopted in this paper follows~\cite{Alessio:2026bdi}.
This is naturally expressed as a double copy of the corresponding QCD Lipatov current~\cite{Lipatov:1976zz,Kuraev:1976ge,Lipatov:1991nf}.

\paragraph{Operator Product Expansion:}
The four-point amplitude $\mm_{2\to2}$ is entirely determined, within the eikonal approximation, by equation~\eqref{eq:amplitude-correlator-2in2}.
Instead, the radiative amplitude $\mm_{2\to3}^\lambda$ we are interested in, is computed through
\begin{equation}\label{eq:amplitude-correlator-2in3}
    i{\rm FT}\big[\mm_{2\to3}^\lambda\big] (s,b) = \Big\la \tilde\Phi_2(b_{2}) \, e^{-\ham\ml_2} \, \hat{a}^\lambda(k) \, e^{-\ham\ml_1} \, \tilde\Phi_1(b_{1}) \Big\ra\,,
\end{equation}
where $\hat{a}^{\lambda}(k)$ is the creation operator of a graviton with momentum $k$ and helicity $\lambda$.
The equation above must be read as follows: evolve the two Wilson lines by applying the evolution operator up to an intermediate rapidity $y_k$, such that $\ml_1 = y_1-y_k\gg1$ and $\ml_2 = y_k-y_2\gg1$; the intermediate rapidity $y_k$ is then identified with the rapidity of the emitted graviton, created by $\hat{a}^\lambda(k)$.
This construction is sketched in Figure~\ref{fig:5pts-from-shocks}, and its practical implementation requires one further ingredient, namely the operator product expansion (OPE) between the creation operator $\hat{a}^\lambda$ and one of the two Wilson lines.
Once $\Phi_a(q)$ is expanded in the reggeon basis, the OPE reads
\begin{equation}\label{eq:OPE}
        \hat{a}^{\lambda}(k) \, W(q) \sim 2\kappa\,\epsilon_{\mu\nu}^{(\lambda)*}(k)\,W(q-k)\,\frac{f^{\mu\nu}(q,k)}{q_\pperp^2} + \mathcal{O}(\kappa^{2}) =
        \hspace{4pt}
        \begin{tikzpicture}[baseline=(base.center),scale=1,transform shape]
            \begin{feynman}
                \vertex[dot] (mid) {};
                \vertex[above=1cm of mid] (init) {\footnotesize$q$};
                \vertex[below=1.1cm of mid] (fin) {\footnotesize$q-k$};
                \vertex[right=1cm of mid] (emission) {\footnotesize$k$};
                \vertex[below=.1cm of mid] (base);
              
                \diagram{
                       (init) -- [edges=photon] (mid) -- [edges=photon] (fin),
                       (mid) -- [edges=graviton] (emission)
                };
                \vertex[dot, style={fill=gray!10, shape=ellipse, minimum width=15pt, minimum height=15pt}, below=1cm of init] (blob) {};
            \end{feynman}
        \end{tikzpicture}
    + \mathcal{O}(\kappa^{2})\,,
\end{equation}
where
\begin{equation}\label{eq:fmunu-def}
       f^{\mu\nu}(q,k) = \left[q_\pperp^2 V^\mu(q,k)\right]\left[q_\pperp^2 V^\nu(q,k)\right]-\frac{q_\pperp^2(q_\pperp-k_\pperp)^2}{k_\pperp^4} \, k_\pperp^\mu \, k_\pperp^\nu\,,
\end{equation}
and
\begin{equation}
       V^\mu(q,k) = \frac{q_\pperp^{\mu}}{q_\pperp^2} - \frac{k_\pperp^{\mu}}{k_\pperp^2}\,.
\end{equation}
The $\mathcal{O}(\kappa)$ term of the OPE expansion~\eqref{eq:OPE} coincides with the insertion of the gravitational Lipatov vertex, namely with the emission of a real graviton in the multi-Regge regime.
Subleading terms involve both loop corrections to the CEV, which are quantum effects, as well as many-reggeon interactions mixing the number of incoming reggeons with that of the outgoing ones.
To give an example, we expect a factor $\sim \kappa^2\,W\,W\,\epsilon_{\mu\nu}^{(\lambda)*}$ to appear at $\mathcal{O}(\kappa^2)$.
One can obtain it either by deriving it through a procedure similar to the one used in QCD~\cite{Caron-Huot:2013fea}, or by matching it to the literature or through other effective theories ({\it e.g.} see the EFT of~\cite{Amati:1993tb} and the recent work of~\cite{Blackstad:2026eoa}).

When working out the OPE with many reggeons, $\hat{a}^\lambda\,W \cdots W$, equation~\eqref{eq:OPE} applies to any pair $(\hat{a}^\lambda,W)$ composing the initial operator, but new terms also appear.
For the purposes of the present work, it suffices to use iterations of $\hat{a}^\lambda\,W$ in equation~\eqref{eq:OPE}, leaving the analysis of the subleading corrections to future work.
We therefore use
\begin{equation}
  \begin{aligned}
    \hat{a}^\lambda(k) W(q_1) W(q_2) \simeq 2\kappa\epsilon_{\mu\nu}^{(\lambda)*}(k) \bigg[
       & W(q_1)W(q_2-k) \frac{f^{\mu\nu}(q_2,k)}{q_{2\pperp}^2} \\
       & + W(q_1-k)W(q_2)\frac{f^{\mu\nu}(q_1,k)}{q_{1\pperp}^2} \bigg] + \mathcal{O}(\kappa^{2})\,.
  \end{aligned}
\end{equation}
The elements introduced above are sufficient to compute the leading logarithmic terms $\mm^{(l,2n),\lambda}_{2\to3}$ in equation~\eqref{eq:m5ll}.
We now justify the ansatz~\eqref{eq:m5ll} based on this construction.
By expanding the Wilson lines in reggeon fields as in~\eqref{eq:wline-expanded} we generate, for each reggeon, an enhancement factor $x^{-1}$.
An additional power of $x^{-2}$ is hidden in the Fourier transform of the amplitude in equation~\eqref{eq:amplitude-correlator-2in3}.
The OPE does not generate any power of $x$, while the evolution Hamiltonian can only control the power of $\log x$ through $\ml_1$ and $\ml_2$, with the constraint that $\ham$ evolves an even (odd) number into another even (odd) number of reggeons.
A simple power counting of $x$ demonstrates that our ansatz~\eqref{eq:m5ll} fits the results at any loop and any order of $x$ at this level of accuracy.


\section{Classical two-loop five-point amplitude at leading logarithmic accuracy}
\label{sec:five-point-amplitude}

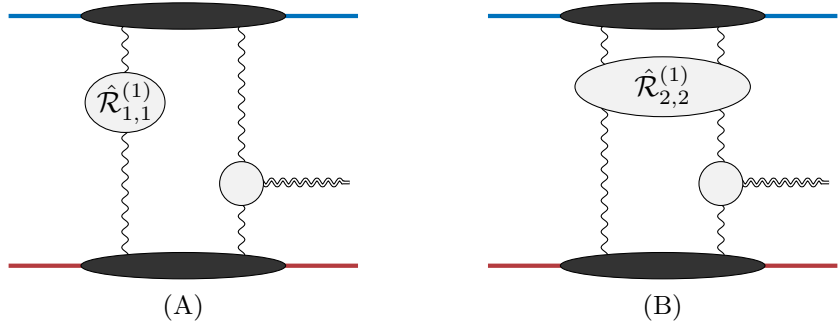
\begin{figure}[t!]
       \vspace{10pt}
       \centering
       \begin{tikzpicture}[scale=1.1,transform shape]
              \begin{feynman}
                     \vertex[dot] (init) {};
                     \vertex[left=2.1cm of init] (in);
                     \vertex[right=2.1cm of init] (out);
                     \vertex[left=.7cm of init] (kul);
                     \vertex[right=.7cm of init] (kur);

                     \vertex[below=3cm of init] (init2);
                     \vertex[left=2.1cm of init2] (in2);
                     \vertex[right=2.1cm of init2] (out2);
                     \vertex[left=.7cm of init2] (kdl);
                     \vertex[right=.7cm of init2] (kdr);

                     \vertex[below=2cm of kur] (ope);
                     \vertex[right=1.3cm of ope] (gout);

                     \vertex[below=3.5cm of init] (init3) {\footnotesize(A)};
              
                     \diagram{
                            {
                                   (in) -- [RoyalBlue!100, ultra thick] (init) -- [RoyalBlue!100, ultra thick] (out),
                                   (in2) -- [cardinal!80, ultra thick] (init2) -- [cardinal!80, ultra thick] (out2),
                                   (kul) -- [edges=photon] (kdl),
                                   (kur) -- [edges=photon] (kdr),
                                   (ope) --[edges=graviton] (gout)
                            }
                     };
                     \vertex[dot, style={fill=black!80, shape=ellipse, minimum width=70pt, minimum height=9pt}, above=0cm of init] {};
                     \vertex[dot, style={fill=black!80, shape=ellipse, minimum width=70pt, minimum height=9pt}, below=3cm of init] {};
                     \vertex[dot, style={fill=gray!10, shape=ellipse, minimum width=15pt, minimum height=15pt}, below=0.68cm of kul] (blob) {\small$\hat{\mathcal{R}}_{1,1}^{(1)}$};
                     \vertex[dot, style={fill=gray!10, shape=ellipse, minimum width=15pt, minimum height=15pt}, below=1.75cm of kur] {};
              \end{feynman}
       \end{tikzpicture}
       \hspace{40pt}
       \begin{tikzpicture}[scale=1.1,transform shape]
              \begin{feynman}
                     \vertex[dot] (init) {};
                     \vertex[left=2.1cm of init] (in);
                     \vertex[right=2.1cm of init] (out);
                     \vertex[left=.7cm of init] (kul);
                     \vertex[right=.7cm of init] (kur);

                     \vertex[below=3cm of init] (init2);
                     \vertex[left=2.1cm of init2] (in2);
                     \vertex[right=2.1cm of init2] (out2);
                     \vertex[left=.7cm of init2] (kdl);
                     \vertex[right=.7cm of init2] (kdr);

                     \vertex[below=2cm of kur] (ope);
                     \vertex[right=1.3cm of ope] (gout);

                     \vertex[below=3.5cm of init] (init3) {\footnotesize(B)};
              
                     \diagram{
                            {
                                   (in) -- [RoyalBlue!100, ultra thick] (init) -- [RoyalBlue!100, ultra thick] (out),
                                   (in2) -- [cardinal!80, ultra thick] (init2) -- [cardinal!80, ultra thick] (out2),
                                   (kul) -- [edges=photon] (kdl),
                                   (kur) -- [edges=photon] (kdr),
                                   (ope) --[edges=graviton] (gout)
                            }
                     };
                     \vertex[dot, style={fill=black!80, shape=ellipse, minimum width=70pt, minimum height=9pt}, above=0cm of init] {};
                     \vertex[dot, style={fill=black!80, shape=ellipse, minimum width=70pt, minimum height=9pt}, below=3cm of init] {};
                     \vertex[dot, style={fill=gray!10, shape=ellipse, minimum width=60pt, minimum height=15pt}, below=0.85cm of init] (blob) {\small$\hat{\mathcal{R}}_{2,2}^{(1)}$};
                     \vertex[dot, style={fill=gray!10, shape=ellipse, minimum width=15pt, minimum height=15pt}, below=1.75cm of kur] {};
              \end{feynman}
       \end{tikzpicture}
       \caption{The figure above is a graphical representation of the amplitudes in equation~\eqref{eq:leading-log-amplitude-2L}, specifically those in the first line, in which the operators $\hat{\mathcal{R}}_{i,j}^{(1)}$ act by evolving the upper Wilson line in rapidity.
       These diagrams are proportional to $\ml_1$. Diagrams such as (A) are obtained by evolving a single reggeon at a time via $\hat{\mathcal{R}}_{1,1}^{(1)}$, whereas diagram (B) evolves, irreducibly, the pair of reggeons emitted by the upper particle.
       Similarly, the second line of equation~\eqref{eq:leading-log-amplitude-2L} is obtained by applying $\tp$, that is, by flipping these diagrams upside down.
       In particular, we see that diagram (B) is nothing but a radiative version of the $H$-diagram~\cite{Amati:1990xe}, in which the graviton is emitted in the MRK via a CEV.}
       \label{fig:two-diags}
\end{figure}

The tree-level amplitude in multi-Regge kinematics~\cite{Raj:2024xsi,Raj:2025hse,Barcaro:2025ifi}, which relies on the tree-level graviton CEV~\cite{Lipatov:1982it}, has recently been re-derived from the shockwave formalism~\cite{Alessio:2026bdi}.
The one-loop amplitude can likewise be derived from the shockwave formalism, but at the leading logarithmic accuracy developed in Sec.~\ref{sec:shockwave-formalism} it yields only terms that are either superclassical or quantum in the classical limit, and are therefore not relevant for the classical study of two-body scattering in the ultrarelativistic limit.
Moreover, the one-loop amplitude from the scattering of two massless particles has recently been computed in~\cite{Blackstad:2026eoa,Rosi:2026zjp}.
As noted in the previous section, reproducing it within our formalism requires extending the OPE by one further order.
We find the first non-trivial classical contribution at two loops, characterised by the following terms
\begin{equation}\label{eq:leading-log-amplitude-2L}
       \begin{aligned}
              \mm_{2\to3}^{(2,2),\lambda} =
              & - \ml_1 \Big\la \Phi_2^{(2)}(q_2) \hat{a}^{\lambda}(k) \Big[ \hat{\mathcal{R}}_{1,1}^{(1)} + \hat{\mathcal{R}}_{2,2}^{(1)} \Big] \Phi_1^{(2)}(q_1) \Big\ra \\
              & - \ml_2 \Big\la \Phi_2^{(2)}(q_2) \Big[ \hat{\mathcal{R}}_{1,1}^{(1)} + \hat{\mathcal{R}}_{2,2}^{(1)} \Big] \hat{a}^{\lambda}(k) \Phi_1^{(2)}(q_1) \Big\ra \,,
       \end{aligned}
\end{equation}
in which we define the rapidity gaps to be
\begin{equation}
       \ml_1 = y_1-y_k \simeq \log\left(\frac{s_{45}}{Q^2}\right)\,, \hspace{25pt} \ml_2 = y_k-y_2 \simeq \log\left(\frac{s_{34}}{Q^2}\right)  \,,
\end{equation}
with $Q$ taken as a reference scale.
Equation~\eqref{eq:leading-log-amplitude-2L} represents the leading logarithmic contribution to the classical part of the two-loop radiative amplitude in the ultrarelativistic limit, and arises from two types of diagrams, depicted in Figure~\ref{fig:two-diags}.
The first type of diagram is obtained by applying $\hat{\mathcal{R}}_{1,1}^{(1)}$ to a two-reggeon state, namely by inserting a one-loop Regge trajectory on one of the two reggeon legs, and yields a diagram that is two-particle reducible (2PR).
The second class of diagrams is instead obtained by evolving the two-reggeon state connectedly, using $\hat{\mathcal{R}}_{2,2}^{(1)}$, and gives a three-particle reducible (3PR) graph.

Moreover, it can be shown explicitly that all the contributions in equation~\eqref{eq:leading-log-amplitude-2L} can equivalently be obtained by implementing the optical theorem for the scattering of massive particles, and by requiring the intermediate, on-shell particles to satisfy the MRK ordering.
This implies that the massive, leading logarithmic five-point, two-loop classical amplitude features a smooth ultrarelativistic limit, namely
\begin{equation}
       i \mathrm{Im}\Big[\mm_{\rm{2\to3},\,massive}^{(2),\lambda}\Big]_{\rm classical,\,LL} \xrightarrow{\rm \hspace{5pt} ultrarelativistic \hspace{5pt}} \; \mm_{2\to3}^{(2,2),\lambda} \,.
\end{equation}
Within this approach, the divergent logarithms $\ml_1$ and $\ml_2$ enter the game as a hard cutoff for the divergent three-particle phase space integrals or through the one-loop Regge trajectory~\cite{Alessio:2026bdi}.
From a technical standpoint, the only tools required to prove that statement are the optical theorem and the MRK approximation of the two- and three-particle phase space, presented in Appendix A of~\cite{Alessio:2026bdi}.\footnote{Unlike the $H$-diagram computation~\cite{Amati:1990xe}, here it is not trivial to obtain the non-log part of the amplitude from the leading logarithmic one by inferring a crossing-symmetric ansatz. The parameter space of a five point process is too large to formulate a unique ansatz for the high-energy behaviour of the full classical amplitude.}
This extends the validity of the results presented in this section to that framework, making them relevant for the study of binary black-hole scattering.

A second comment can be made by distinguishing the diagrams in the first line of equation~\eqref{eq:leading-log-amplitude-2L} from those in the second.
In the high-energy limit of scattering amplitudes there exists a symmetry, termed {\it target-projectile symmetry}, which exchanges the two bodies, $\Phi_1 \leftrightarrow \Phi_2$ leaving the result untouched.
This is realised as the exchange $P_1 \leftrightarrow P_2$ and $P_5 \leftrightarrow P_3$ and, since the evolution Hamiltonian is self-adjoint, the contributions in the second line can be obtained by applying target-projectile operation to those in the first, or vice versa.
This has been verified explicitly for the present computation.
The extension of target-projectile symmetry to the massive case has been discussed {\it e.g.} in the papers on the one-loop gravitational waveform, see {\it e.g.}~\cite{Georgoudis:2023lgf}.
In order to make efficient use of this symmetry, we denote the target-projectile exchange operation by $\tp$, and label $\ma_{\rm 2PR}$ and $\ma_{\rm 3PR}$ respectively the contributions coming from $\hat{\mathcal{R}}_{1,1}^{(1)}$ and $\hat{\mathcal{R}}_{2,2}^{(1)}$, so that
\begin{equation}\label{eq:M5-21-master}
      \mm_{2\to3}^{(2,2),\lambda} = \ml_2\left(\ma_{\rm 2PR}^\lambda+\ma_{\rm 3PR}^\lambda\right) + \ml_1\left(\tp*\ma_{\rm 2PR}^\lambda+\tp*\ma_{\rm 3PR}^\lambda\right)\,.
\end{equation}
An explicit use of the shockwave formalism -- evolution operator and OPE -- gives
\begin{align}
       & \ma_{\rm 3PR}^\lambda = \frac{i\kappa^7 s^3}{4\pi x^6} \, \epsilon_{\mu\nu}^{(\lambda)*}(k) \, \mathcal{F}_{\rm 3PR}^{\mu\nu} \,, \label{eq:a3pr-def} \\
       & \mathcal{F}_{\rm 3PR}^{\mu\nu} = \int_{\ell_{1\pperp},\ell_{2\pperp}}\frac{\mathcal{H}_{\mathrm{GR}}(\ell_2-\ell_1,\ell_2;-q_2)\;f^{\mu\nu}(\ell_2-\ell_1+k,k)}{\ell_{2\pperp}^2(q_2+\ell_{2})_\pperp^2(\ell_{1}-\ell_{2})_\pperp^2(q_{2}+\ell_{2}-\ell_{1})_\pperp^2 (\ell_{2}-\ell_{1}+k)_\pperp^2} \,, \label{eq:f3pr-def}
\end{align}
for the 3PR term, and
\begin{align}
       & \ma_{\rm 2PR}^\lambda = \frac{i\kappa^7 s^3}{4\pi x^6} \, \epsilon_{\mu\nu}^{(\lambda)*}(k) \, \mathcal{F}_{\rm 2PR}^{\mu\nu} \,, \label{eq:a2pr-def} \\
       & \mathcal{F}_{\rm 2PR}^{\mu\nu} = -\int_{\ell_{1\pperp},\ell_{2\pperp}} \frac{(d+1)f^{\mu\nu}(\ell_2+k,k)}{\ell_{2\pperp}^2 (q_2+\ell_2)_\pperp^2 (\ell_2+k)_\pperp^2} \left[\frac{(\ell_{2\pperp}^2)^2}{\ell_{1\pperp}^2(\ell_1-\ell_2)_\pperp^2} + \frac{((q_2+\ell_2)_\pperp^2)^2}{\ell_{1\pperp}^2(q_2+\ell_2-\ell_1)_\pperp^2}\right], \label{eq:f2pr-def}
\end{align}
for the 2PR one.
Then we use momentum conservation $k=q_1+q_2$ to express the integrals in term of $q_1$ and $q_2$ only.
Given the symmetry of the prefactors in equations~\eqref{eq:a3pr-def} and~\eqref{eq:a2pr-def} under target-projectile operation, $\tp$ reduces to the exchange $q_{1\pperp} \leftrightarrow q_{2\pperp}$ in the transverse plane.
This follows from the factorisation of the light-cone and transverse dynamics at high energies, which reduces the integrals of a five-point process in $D$ dimensions to the computation of the integrals $\mathcal{F}^{\mu\nu}_{i}$, describing an off-shell three-point process $q_{1\pperp} + q_{2\pperp} \to k_\pperp$ in $d=D-2$ dimensions.
It is therefore clear that the operation $\tp$, acting on the amplitude, is nothing but an element of the group $S_3$ acting on the integrals $\mathcal{F}^{\mu\nu}_i$, permuting the external legs $\{q_{1\pperp},q_{2\pperp},k_\pperp\}$.
These transformations were first introduced in~\cite{Chavez:2012kn} in the context of the three-point massless integrals, and are reviewed in Appendix~\ref{app:permutation-S3}.
In the present case, it suffices to use a single element of $S_3$,
\begin{equation}
       \tp*\ma_i^\lambda \quad \longleftrightarrow \quad g_1*\mathcal{F}_i^{\mu\nu} \,,
\end{equation}
namely $g_1 = \{q_{1\pperp} \leftrightarrow q_{2\pperp} \} = \{ z \leftrightarrow 1-\zb \}$.

\subsection{Computational details}\label{sec:computation}
The computation of the integrals $\mathcal{F}_i^{\mu\nu}$ is divided into the following steps:
$(i)$ we choose a tensor basis in which to express the function $f^{\mu\nu}$ as a combination of external momenta, so as to be left with purely scalar integrals;
$(ii)$ we implement the integration-by-parts identities (IBPs), reducing the computation to 11 master integrals (MIs);
$(iii)$ finally, using the permutation group $S_3$, we reduce the computation further, to just five MIs, whose solutions are given in terms of single-valued polylogarithms (SVPLs).

We note that the functions $\mathcal{F}_i^{\mu\nu}$, built entirely out of $(2-2\veps)$-dimensional transverse momenta, are multiplied by the polarisation tensor $\epsilon_{\mu\nu}^{(\lambda)*}$, which lives in $2$ dimensions.\footnote{Regarded as a tensor in $2-2\varepsilon$ dimensions, this vanishes whenever one of its indices lies in the $-2\varepsilon$ dimensions.}
We therefore introduce a decomposition of the $(2-2\veps)$-dimensional space as the sum of projectors along the two directions defined by $q_{1\pperp}$ and $q_{2\pperp}$, together with a projector onto the remaining $-2\veps$ dimensions, which nonetheless vanishes when contracted with the polarisation tensor.
Using the convention $\ell^i = (\vec{\ell}\,)^i$ for latin indices, we write
\begin{equation}\label{eq:projectors}
    \delta_{(2-2\epsilon)}^{ij} = \Pi_1^{ij} + \Pi_2^{ij} + \Pi_{-2\veps}^{ij}\,
\end{equation}
where the projectors satisfy $\Pi_a^{ik}\Pi_b^{kj} = \delta_{ab} \, \Pi_a^{ij}$, and are expressed as $\Pi_{a}^{ij} = q_{a}^i \check{q}_{a}^j$, with
\begin{gather}
    \label{eq:check-variables}
    \check{q}_1^i = \frac{q_{2\pperp}^2 q_{1}^i - (q_{1\pperp}\cdot q_{2\pperp}) q_{2}^i}{(q_{1\pperp}\cdot q_{2\pperp})^2-q_{1\pperp}^2q_{2\pperp}^2} \,,
    \quad\quad
    \check{q}_2^i = \frac{q_{1\pperp}^2 q_{2}^i - (q_{1\pperp}\cdot q_{2\pperp}) q_{1}^i}{(q_{1\pperp}\cdot q_{2\pperp})^2-q_{1\pperp}^2q_{2\pperp}^2} \,.
\end{gather}
for $a=1,2$. We note the presence of spurious poles in the denominators of the dual variables $\check{q}_a$, which cancel by the end of the computation.
These poles correspond to the zeros of the function $\lambda(1,u,v)$, and thus to the case in which the three momenta are collinear in the transverse space.
Given the decomposition~\eqref{eq:projectors} and the polarisation tensor $\epsilon_{\mu\nu}^{(\lambda)}$, which is symmetric $\epsilon_{\mu\nu}^{(\lambda)} = \epsilon_{\nu\mu}^{(\lambda)}$, we may define three tensor structures,
\begin{subequations}
       \begin{align}
              & T_{1}^{\mu\nu} = q_{1\pperp}^\mu q_{1\pperp}^\nu \,,\\
              & T_{2}^{\mu\nu} = q_{2\pperp}^\mu q_{2\pperp}^\nu \,,\\
              & T_{3}^{\mu\nu} = \frac{1}{2}\left(q_{1\pperp}^\mu q_{2\pperp}^\nu + q_{2\pperp}^\mu q_{1\pperp}^\nu \right)\,,
       \end{align}
\end{subequations}
such that any tensor integral can be expressed as a linear combination of these,
\begin{equation}
       \mathcal{F}_i^{\mu\nu} = \sum_{A=1,2,3} \mathcal{F}_i^{(A)} \cdot T_{A}^{\mu\nu} \,.
\end{equation}
We then implement the IBPs for the scalar integrals $\mathcal{F}_i^{(A)}$, using the package $\texttt{LiteRed}$~\cite{Lee:2013mka} and the following basis of denominators,
\begin{equation}\label{eq:ibp-basis}
       \Big\{ \ell_{1\pperp}^2 , \ell_{2\pperp}^2 , (\ell_{1}-\ell_{2})_\pperp^2 , (\ell_{2}+q_{1}+q_{2})_\pperp^2 , (\ell_{2}+q_{2})_\pperp^2 , (\ell_{1}-\ell_{2}-q_{2})_\pperp^2 , (\ell_{1}-\ell_{2}-q_{1}-q_{2})_\pperp^2 \Big\}.
\end{equation}
The resulting list of 11 master integrals, and its reduction to just 5 independent integrals, is discussed in Appendix~\ref{app:integrals}.

A brief comment on the nature of the integrals obtained is in order.
Those integrals have been studied in detail in $D=4-2\varepsilon$~\cite{Usyukina:1994iw,Birthwright:2004kk,Chavez:2012kn}; in particular, once expressed in the variables $(z,\zb)$ defined in eq.~\eqref{eq:zbarz-from-uv}, they involve only single-valued (SV) functions~\cite{Chavez:2012kn} in the physical region, in which $z$ and $\zb$ are complex conjugates.
The same integrals in $d$ dimensions have been studied in the context of the shockwave formalism in QCD~\cite{Abreu:2024xoh,Buccioni:2024gzo}.
Given their simplicity, we computed them in two independent ways as a cross-check of our results.
The first is by explicit calculation using Feynman parameters, and subsequently integrating the result using \texttt{PolyLogTools}~\cite{Duhr:2019tlz}.
The second method makes use of the results obtained in~\cite{Chavez:2012kn} in the Euclidean region - which is the one of interest here - together with dimensional shift relations~\cite{Weinzierl:2022eaz,Derkachov:1990osa,Lee:2010wea,Tarasov:1996br}.
The derivation and results for the individual MIs are presented in Appendix~\ref{app:integrals}.

\subsection{Results}\label{sec:results}
Using the eleven MIs, IBP identities and tensor reduction, we obtain the following results for the two positive-helicity contributions depicted in Figure~\ref{fig:two-diags},
\begin{equation}\label{eq:a3pr-sol}
    \begin{aligned}
        \ma^\oplus_{\rm 3PR} = & - \frac{i\kappa^7 s^3}{64\pi^3 x^6} \left(\frac{4\pi\mu^2}{Q^2}\right)^{2\veps} \frac{1}{z}\; \bigg\{ \frac{2z-z^2+z\zb}{2\veps^2} \\
        & - \frac{z \zb \log \left(z \zb\right)+(1-z)(z-\zb)-(1-z)^2 \log\left((1-z)(1-\zb)\right)}{\veps} \\
        & + \frac{1-z}{z-\zb} \left(3z\zb+z-3\zb-\zb^2\right)D_2(z,\bar z) + z\zb\log^2(z\zb) - 2(1-z)\zb\log(z\zb) \\
        & - \frac{1}{2}(1-z)(\zb-2z+1)\log^2\left((1-z)(1-\zb)\right)- 2(1-z)^2\log\left((1-z)(1-\zb)\right) \\
        & - \frac{1}{2}(1-z)(1-\zb)\log\left(z\zb\right)\log\left((1-z)(1-\zb)\right) + \frac{\zeta_2}{2}(z^2-z\zb-2z) + z+\zb + \mathcal{O}(\veps) \bigg\},
    \end{aligned}
\end{equation}
and
\begin{equation}\label{eq:a2pr-sol}
       \begin{aligned}
              \ma^\oplus_{\rm 2PR} = & - \frac{i\kappa^7 s^3}{64\pi^3 x^6} \left(\frac{4\pi\mu^2}{Q^2}\right)^{2\veps} \frac{1}{z}\; \Bigg\{ \frac{3 (z^2+2\zb-z\zb)}{2\veps^2} \\
              & - \frac{z^2 - 3z - \zb - z\zb - 3(z-\zb)(1-z) \log((1-z)(1-\zb)) + 3\zb\log(z\zb)}{\veps} \\
              & + 3(z-\zb)D_2(z,\zb) - 3(z-\zb)(1-z)\log^2((1-z)(1-\zb)) \\
              & + \frac{3}{2}\zb\log^2(z\zb) - \frac{3}{2}(z-\zb)\log(z\zb)\log((1-z)(1-\zb)) - (3z+1)\zb\log(z\zb) \\
              & - (1-z)(2z+\zb-3)\log((1-z)(1-\zb)) + \frac{3}{2}\zeta_2(z\zb-2\zb-z^2) + 7(z+\zb) + \mathcal{O}(\veps)
              \Bigg\}\,,
       \end{aligned}
\end{equation}
whose sum gives leading-logarithmic part of the two-loop amplitude as in eq.~\eqref{eq:M5-21-master}. We stress that the poles in $z=\zb$, which is the zero of $\lambda(1,u,v)$, cancel from the final result.

\subsection{Leading soft factorisation}\label{sec:tests}
To verify the correctness of our results in equations~\eqref{eq:a3pr-sol} and~\eqref{eq:a2pr-sol}, we perform a standard test, namely the leading soft factorisation theorem by Weinberg~\cite{Weinberg:1964ew,Weinberg:1965nx}.
In the limit in which the graviton emitted is soft compared to all the other scales involved, a gravitational amplitude factorises as the product of a hard-scattering part and a soft current, whose leading order is divergent and inversely proportional to the emitted frequency.
This current originates from emissions off external legs.
Therefore, it is universal both in the loop expansion and in the number of particles involved in the process.
Expressing the graviton momentum as $k^\mu = \omega n^\mu$, where $\omega$ is its frequency and $n^\mu = (2c^2_{\theta/2},2s^2_{\theta/2},s_\theta c_\phi,s_\theta s_\phi)$ its direction, with the shorthands $s_\alpha = \sin\alpha$ and $c_\alpha = \cos\alpha$, in the limit $\omega\to0$ one obtains
\begin{equation}\label{eq:leading-soft-thm}
       \mm_{2\to3}^{(2,2),\lambda} = \mm_{2\to2}^{\text{\tiny LL}} \, J_{[1/\omega]}^{\mu\nu} \, \epsilon^{(\lambda)*}_{\mu\nu} + \mathcal{O}(\omega^0,\log\omega)\,,
\end{equation}
where $\mm_{2\to2}^{\text{\tiny LL}}$ is the corresponding classical four-point, two-loop amplitude in the {\it leading logarithmic} approximation, computed in~\cite{Amati:1990xe}, and $J^{\mu\nu}_{[1/\omega]}$ is the Weinberg current, given by
%
\begin{equation}\label{eq:weinberg-current}
       \begin{aligned}
              J^{\mu\nu}_{[1/\omega]} 
              & = \frac{\kappa}{\omega} \sum_{i=1,2,3,5} \frac{P_i^\mu P_i^\nu}{n\cdot P_i} \\
              & = \frac{\kappa}{\omega} \bigg[ q\!\cdot\! n \left( \frac{p_1^{\mu}p_1^{\nu}}{(p_1\!\cdot\! n)^2} - \frac{p_2^{\mu}p_2^{\nu}}{(p_2\!\cdot\! n)^2} \right)
                     - \frac{p_1^{\mu}q^{\nu}+p_1^{\nu}q^{\mu}}{p_1\!\cdot\! n} + \frac{p_2^{\mu}q^{\nu}+p_2^{\nu}q^{\mu}}{p_2\!\cdot\! n} \bigg],
       \end{aligned}
\end{equation}
where $q \simeq q_1 \simeq -q_2$ in the $\omega \to 0$ limit.
Contracting with the light-cone polarisations~\eqref{eq:grap-pols}, we get an extremely simple form of the Weinberg current
%
%
\begin{equation}
    J_{[1/\omega]}^\oplus = J^{\mu\nu}_{[1/\omega]} \, \epsilon^{\oplus\,*}_{\mu\nu} = - \frac{2\kappa}{\omega n_\pperp^2} \left[ \frac{n\cdot q}{n_\pperp^2} (\zeta_n^*)^2 - \zeta_n^* \zeta_q^* \right] + \mathcal{O}(x) = - \kappa \left(z-\zb\right) + \mathcal{O}(x) \,.
\end{equation}
As for the classical radiative amplitude, we likewise split the two-loop conservative classical amplitude $\mm_{2\to2}^{\text{\tiny LL}}$ into a contribution obtained through evolution via $\hat{\mathcal{R}}_{1,1}^{(1)}$ and one via $\hat{\mathcal{R}}_{2,2}^{(1)}$, which we denote, respectively, as
\begin{equation}\label{eq:amplitude-4-sectors}
       \mm_{2\to2}^{(2),\text{\tiny LL}} = \left[\ma_{{2\to2},{\rm 2PR}}^{(2)} + \ma_{{2\to2},{\rm 3PR}}^{(2)}\right]\ml \,,
\end{equation}
with $\ml = \log(-s/t)$ as in equation~\eqref{eq:amplitude-correlator-2in2}.
These two contributions can be obtained from the shockwave formalism, following the prescription in equation~\eqref{eq:amplitude-correlator-2in2}, and are graphically identical to the diagrams in Figure~\ref{fig:two-diags} up to the OPE insertion.
The first of these gives a 2PR contribution, obtained by sewing together a one-loop quantum amplitude involving a Regge trajectory with a classical tree-level reggeon exchange, whereas the term $\ma_{{2\to2},{\rm 3PR}}^{(2)}$ is the well-known $H$-diagram, first computed by Amati, Ciafaloni and Veneziano~\cite{Amati:1990xe} and reproduced in~\cite{Alessio:2026bdi,Alessio:2025isu,Raj:2025hse,DiVecchia:2020ymx}.
Indeed, taking the soft limit, we recover exactly the expected conservative amplitudes,
\begin{align}
       \label{eq:H-diagram-soft}
       & \ma_{\rm3PR}^\oplus\ml_2 + \left(\tp*\ma_{\rm3PR}^\oplus\right)\ml_1 \simeq \, \underbrace{\frac{i\kappa^6 s^3\ml}{128\pi^3x^6} \left(\frac{4\pi\mu^2}{-t}\right)^{2\veps} \left(-\frac{1}{\veps^2} + \frac{2}{\veps} + \zeta_2 + \mathcal{O}(\veps)\right)}_{\ma_{{2\to2},{\rm 3PR}}^{(2)}\ml} \; J_{[1/\omega]}^\oplus \,, \\[5pt]
       \label{eq:regge-2L-diagram-soft}
       & \ma_{\rm2PR}^\oplus\ml_2 + \left(\tp*\ma_{\rm2PR}^\oplus\right)\ml_1 \simeq \underbrace{ \frac{i\kappa^6 s^3 \ml}{128\pi^3x^6} \left(\frac{4\pi\mu^2}{-t}\right)^{2\veps} \left(\frac{3}{\veps^2} - \frac{2}{\veps} - 3\zeta_2 + \mathcal{O}(\veps)\right)}_{\ma_{{2\to2},{\rm 2PR}}^{(2)}\ml} \; J_{[1/\omega]}^\oplus \,,
\end{align}
up to corrections $\mathcal{O}(\omega^0,x)$.
Here we used the fact that $t_1 \simeq t_2 \simeq t$ in the soft limit, and hence $\ml_1 + \ml_2 \simeq \ml$.


\section{Conclusions}\label{sec:conclusions}

We have computed the classical part of the two-loop five-point amplitude at leading logarithmic accuracy for the emission of a graviton in the collision of two massless scalar fields in Einstein--Hilbert gravity, in multi-Regge kinematics.
The amplitude follows from the shockwave formalism for massless particles presented in~\cite{Alessio:2026bdi}, which is in turn formulated by applying double-copy relations to the building blocks of high-energy QCD, namely the tree-level Lipatov CEV, the BFKL integral kernel and the one-loop Regge trajectory of the graviton.
Within this formalism the two colliding particles are described by Wilson lines, which are expanded in reggeon fields and then evolved through a rapidity evolution equation.
Once the graviton has been emitted through the operator product expansion~\eqref{eq:OPE} between the creation operator and a reggeon field, the amplitude is obtained as the time-ordered correlator of the remaining reggeons.
Each application of the evolution Hamiltonian $\ham$ generates one large rapidity logarithm and carries at least one factor of $\kappa^2$.
Nevertheless, the structure of $\ham$ is such that the first classical logarithm only appears at two loops, so that the $\mathcal{O}(\kappa^5)$ order can be neglected at our accuracy.

The amplitude presented in equation~\eqref{eq:leading-log-amplitude-2L} features two distinct structures, one involving the one-loop graviton Regge trajectory and the other the gravitational BFKL kernel.
The first gives a two-particle reducible diagram, amounting to the convolution, in the transverse space, of a quantum correction with a classical exchange~\cite{DiVecchia:2023frv}.
On the other side the second structure generates three-particle reducible diagrams, which represent the radiative corrections to the $H$-diagram in which the emission is constrained in the MRK region.
High-energy factorisation reduces their computation to two-loop three-point integrals in $d=2-2\veps$ transverse dimensions, so that the transverse dependence is carried, at $\mathcal{O}(\veps^0)$, by single-valued polylogarithms of weight at most two in the single complex variable $z$, while the longitudinal invariants enter only through the two rapidity gaps factored out of the integrals.

We point out that the same result can be obtained as the imaginary part of the classical $2\to3$ amplitude for massive scalar fields, taken in the ultrarelativistic limit and with the strict multi-Regge constraints imposed on the $S$-channel cut.
This strengthens the correspondence established in~\cite{Alessio:2026bdi} between the leading logarithmic classical amplitudes describing the scattering of massless and of massive particles, and extends it to radiative amplitudes.
We nevertheless expect such a universality to break down beyond the leading logarithmic approximation.

Our accuracy is strictly leading logarithmic in $x$ and the size of the five-point parameter space makes it impractical to reconstruct the subleading logarithms from an ansatz, as was done for the elastic case in~\cite{Amati:1990xe,DiVecchia:2021bdo}.
At one loop the same construction yields only superclassical or quantum terms, and the classical one-loop amplitude, calculated in~\cite{Blackstad:2026eoa,Rosi:2026zjp}, would be recovered only by extending the operator product expansion by one order in $\kappa$.
Nevertheless, both the 2PR and the 3PR contributions to the classical leading logarithmic amplitude give null contribution to the emitted waveform, once the KMOC-Cristofoli-Gonzo formalism~\cite{Kosower:2018adc,Cristofoli:2021vyo} is applied.
Therefore we conclude that a next-to-leading treatment is necessary and, at this level, we expect a breakdown of the universality of the ultrarelativistic limit on the Post-Minkowskian terms of the waveform.
As a consequence, the corrections to the shockwave formalism will be valid for massless-massless scattering, while different methods are required to describe the emitted waveform of massive-massive scattering at high velocity.


\section*{Acknowledgments}
We thank Stefano De Angelis and Carlo Heissenberg for useful discussions. We are also grateful to Francesco Alessio for his collaboration in the early stages of this project.

\appendix

\section{Integrals}\label{app:integrals}
Using the basis of~\eqref{eq:ibp-basis} to reduce the two-loop, three-point massless integrals via IBPs, we get a set of 11 MIs
\begin{subequations}\label{eq:mis}
    \begin{align}
       & \mi{a} =-\mi{1,0,1,1,0,0,0} = \mathcal{B}_{1,1}^{(2)}(q_1+q_2) \,, \label{mi1} \\
       & \mi{b} =-\mi{1,0,0,0,1,0,1} = \mathcal{B}_{1,1}^{(2)}(q_1) \,, \label{mi2} \\
       & \mi{c} = \mi{0,0,1,1,1,1,0} = \mathcal{B}_{1,1}(q_1)\,\mathcal{B}_{1,1}(q_2) \,, \label{mi3} \\
       & \mi{d} = \mi{1,0,1,0,1,0,1} = \left[\mathcal{T}\star\mathcal{B}_{1,1}\right] (q_1,q_2) \,, \label{mi4} \\
       & \mi{e} = \mi{1,0,1,1,1,0,0} = \left[\mathcal{T}\star\mathcal{B}_{1,1}\right] (q_1+q_2,q_2) \,, \label{mi5} \\
       & \mi{f} =-\mi{1,0,1,0,1,0,0} = \mathcal{B}_{1,1}^{(2)}(q_2) \,, \label{mi6} \\
       & \mi{g} = \mi{0,1,1,0,1,0,1} = \mathcal{B}_{1,1}(q_2)\,\mathcal{B}_{1,1}(q_1+q_2) \,, \label{mi7} \\
       & \mi{h} =-\mi{2,0,1,0,1,0,1} = \left[\mathcal{T}\star\mathcal{B}_{1,2}\right] (q_1,q_2) \,, \label{mi8} \\
       & \mi{i} =-\mi{0,1,1,0,1,1,1} = \mathcal{B}_{1,1}(q_2) \mathcal{T}(q_1,q_2) \,, \label{mi9} \\
       & \mi{j} =-\mi{2,0,1,1,1,0,0} = \left[\mathcal{T}\star\mathcal{B}_{1,2}\right] (q_1+q_2,q_2) \,, \label{mi10} \\
       & \mi{k} = \mi{0,1,1,0,1,1,0} = \left[\mathcal{B}_{1,1}(q_2)\right]^2 \,. \label{mi11} 
    \end{align}
\end{subequations}
The relative signs between the definitions of $\mathcal{I}_{n_1,\cdots,n_7}$ and the $\mathcal{I}_{a,\dots,k}$ are due to the choice of the basis~\eqref{eq:ibp-basis}, where all the denominators are negative.

In equations~\eqref{eq:mis} we have defined two families of integrals, together with a convolution operator $\star$ acting between Feynman integrals, and explained below.
The $\mathcal{B}_{i,j}$ are massless bubble integrals~\cite{Kotikov:2018wxe},
\begin{equation}\label{eq:generalised-bubbles}
       \begin{gathered}
           \mathcal{B}_{a_1,a_2}(q) = \int_{\ell_\pperp} [-(q_\pperp - \ell_\pperp)^2]^{-a_1}[-\ell_\pperp^2]^{-a_2} = \left(\frac{4\pi\mu^2e^{\geps}}{-q_\pperp^2}\right)^\veps \, \frac{B_{a_1,a_2}}{4\pi} \, (-q_\pperp^2)^{1-a_1-a_2}, \\
           B_{a_1,a_2} = \frac{F\left(a_1\right) F\left(a_2\right)}{F\left(a_1+a_2-\frac{d}{2}\right)}, \qquad F(a) = \frac{\Gamma\left(\frac{d}{2}-a\right)}{\Gamma\left(a\right)},
       \end{gathered}
\end{equation}
while the $\mathcal{B}_{1,1}^{(n)}$ are iterated bubbles with seed bubble $\mathcal{B}_{1,1}$,
\begin{equation}
    \mathcal{B}_{1,1}^{\,(n)}(q) = \left(\mu^2 e^{\geps}\right)^{n\veps} \int \left[\prod_{i=1}^{n+1} \frac{\hat{d}^d\ell_{i\pperp}}{-\ell_{i\pperp}^2} \right] \hat{\delta}^{(d)}\bigg(q_\pperp - \sum_{j=1}^{n+1} \ell_{j\pperp} \bigg)
\end{equation}
whose result, for $d=2-2\veps$ and with $b_n = 1+\veps(n-1)$, is
\begin{equation}\label{eq:iterated-bubbles}
    \mathcal{B}_{1,1}^{\,(n)}(q) = \left(\frac{4\pi\mu^2 e^{\geps}}{-q_\pperp^2}\right)^{n\veps} \frac{\Gamma\left(-\veps\right)^{n+1}}{(4\pi)^n\,F(b_{n+1})\,(-q_\pperp^2)}\,.
\end{equation}
The one-loop triangles are defined with the following sign conventions
%
\begin{equation}\label{eq:tri1L}
    \mathcal{T}(v_1,v_2) = \int_{\ell_\perp} \frac{1}{\big[-\ell_\pperp^2\big]\,\big[-(\ell + v_{1})_\pperp^2\big]\big[-(\ell-\delta_{12}v_{2})_\pperp^2\big]}\,,
\end{equation}
in which the sign $\delta_{12}$ is minus if the two vectors $v_1$ and $v_2$ are both ingoing or both outgoing, and plus otherwise.
The symbol $\star$ is a convolution between Feynman integrals.
To give an example, the integral~\eqref{mi4}, after loop-momentum shifts, reads
\begin{equation}
    \mi{d} = \int_{\ell_{1\pperp},\ell_{2\pperp}} \frac{1}{\ell_{2\pperp}^2\,(\ell_{2\pperp}+\ell_{1\pperp})^2\,(\ell_{1\pperp}+q_{2\pperp})^2\,(\ell_{1\pperp}-q_{1\pperp})^2} = \int_{\ell_{1\pperp}} \frac{\mathcal{B}_{1,1}(\ell_1)}{(\ell_{1\pperp}+q_{2\pperp})^2\,(\ell_{1\pperp}-q_{1\pperp})^2}\,.
\end{equation}
In this case, the convolution symbol thus means ``{\it replace a propagator with a bubble}''.
Similarly, iterated bubbles are defined recursively,
\begin{equation}
       \mathcal{B}_{1,1}^{(n)}(q) = \left[\mathcal{B}_{1,1}^{(n-1)}\star\mathcal{B}_{1,1}\right](q)\,.
\end{equation}
From the above definitions it immediately follows that the integrals $\mi{a,b,c,f,g,k}$ are trivially solved using equation~\eqref{eq:iterated-bubbles}, while the remaining MIs are the one-loop triangle $\mathcal{T}(q_1,q_2)$ appearing in $\mi{i}$, and the two-loop convolutions $\mi{d}$ and $\mi{h}$.
The integrals $\mi{e}$ and $\mi{j}$ are in turn obtained from $\mi{d}$ and $\mi{h}$, respectively, by permutation of the external legs,
\begin{equation}\label{eq:mi-from-S3}
    \mi{e} = g_2 * \mi{d}\,,\quad \text{and} \quad \mi{j} = g_2 * \mi{h}\,,
\end{equation}
where the elements $g_a$ are defined in Appendix~\ref{app:permutation-S3}.
The one-loop triangle-times-bubble $\mi{i}$ is obtained using Feynman parameters,
\begin{equation}\label{eq:Ii-MI}
\begin{aligned}
    & \mi{i} = \left(\frac{4\pi\mu^2}{Q^2}\right)^{2\veps} \frac{1}{16 \pi^2 Q^6|z|^2|1-z|^4} \bigg\{ \frac{4z\zb-2\zb-2z+4}{\veps^2} \\
    & \qquad\quad + \frac{1}{\veps}\Big[ 2 \left(\zb+z-2\right) \log \left(z \zb\right)-4 \left(z \zb+1\right) \log\left((1-z)(1-\zb)\right) \Big] + 4(z-\zb)D_2(z,\bar z) \\
    & \qquad\quad - 2\zeta_2 (2z\zb-\zb-z+2) + 2 (z+1) \left(\zb+1\right) \log^2\left((1-z) \left(1-\zb\right)\right) \\
    &\qquad\quad -\left(\zb+z-2\right) \log ^2\left(z \zb\right) -2\left(\zb+z-2\right) \log \left(z
   \zb\right) \log\left((1-z)(1-\zb)\right) + \mathcal{O}(\veps) \bigg\}
\end{aligned}
\end{equation}
where the first non-trivial, weight-two, SVPL, namely the Bloch-Wigner dilogarithm $D_2(z,\bar z)$, occurs
\begin{equation}\label{eq:D2-definition}
    D_2(z,\bar z) = \mathrm{Li}_2(z) - \mathrm{Li}_2(\zb) + \frac{1}{2}\log(z\zb) \log\left(\frac{1-z}{1-\zb}\right) \,.
\end{equation}
It satisfies (see {\it e.g.}~\cite{Chavez:2012kn,Buccioni:2024gzo}) $D_2(1-\zb,1-z) = D_2(z,\zb)$ and $D_2(\zb,z) = D_2(1/z,1/\zb) = -D_2(z,\zb)$, implying that $g_1\ast D_2(z,\zb) = g_2\ast D_2(z,\zb) = D_2(z,\zb)$.
The convoluted triangle-bubble integrals $\mi{d}$ and $\mi{h}$ are, respectively,
\begin{equation}\label{eq:Id-MI}
\begin{aligned}
    & \mi{d} = \left(\frac{4\pi\mu^2}{Q^2}\right)^{2\veps} \frac{1}{16 \pi^2 Q^4|z|^2|1-z|^2} \bigg\{ \frac{4z\zb-2\zb-2 z+3}{\veps^2} \\
    & \qquad\quad - \frac{2}{\veps} \Big[ \left(z \zb+\zb+z\right) \log \left((1-z)(1-\zb)\right) +\left(z\zb-2z-2\zb+3\right) \log\left(z\zb\right) \Big]  \\
    & \qquad\quad + 6(z-\zb)D_2(z,\bar z) + (z+\zb-2z\zb)\log(z\zb)\log((1-z)(1-\zb))  \\
    & \qquad\quad + 2(z\zb-2z-2\zb+3)\log^2(z\zb) + 2(z\zb+z+\zb)\log^2((1-z)(1-\zb)) \\
    & \qquad\quad - \zeta_2(4z\zb-2z-2\zb+3) + \mathcal{O}(\veps) \bigg\}\,,
\end{aligned}
\end{equation}
\begin{equation}\label{eq:Ih-MI}
\begin{aligned}
    & \mi{h} = \left(\frac{4\pi\mu^2}{Q^2}\right)^{2\veps} \frac{1}{16 \pi^2 Q^6 |z|^4|1-z|^4} \bigg\{\frac{4 z^2 \zb^2-4 z^2 \zb-4 z \zb^2+12 z \zb+2 \zb^2+2 z^2-6 z-6 \zb+3}{\veps^2} \\
    & \qquad\quad + \frac{1}{\veps} \Big[8z^2\zb^2 - 8z^2\zb - 8z\zb^2 + 38z\zb + 6\zb^2 + 6z^2-21z - 21\zb + 12
    - \log\left(z\zb\right) \big(2 z^2 \zb^2 \\
    & \qquad\quad -8 z^2 \zb-8 z \zb^2+24 z \zb+4 \zb^2-12 \zb+4 z^2-12 z+6\big)
    - \log((1-z)(1-\zb))\big(2 z^2 \zb^2 \\
    & \qquad\quad +4 z^2 \zb+4 z \zb^2-2 \zb^2-2 z^2\big) \Big]
    + 6 \left(2z^2\zb - 2z\zb^2 - z^2 + \zb^2\right) D_2(z,\bar z) \\
    & \qquad\quad - \zeta_2 \left(4 z^2 \zb^2-4 z^2 \zb-4 z \zb^2+12 z \zb+2 \zb^2-6 \zb+2 z^2-6 z+3\right) \\
    & \qquad\quad + \log^2(z\zb)\left(2 z^2 \zb^2-8 z^2 \zb-8 z \zb^2+24 z \zb+4 \zb^2-12 \zb+4 z^2-12 z+6\right) \\
    & \qquad\quad + \log^2\left((1-z)(1-\zb)\right)\left(2 z^2 \zb^2+4 z^2 \zb+4 z \zb^2-2 \zb^2-2 z^2\right) \\
    & \qquad\quad - \log(z\zb)\log\left((1-z)(1-\zb)\right)\left(2 z^2 \zb^2-2 z^2 \zb-2 z \zb^2+\zb^2+z^2\right) \\
    & \qquad\quad - \log(z\zb)\big(4 z^2 \zb^2-22 z^2 \zb-22 z \zb^2+76 z \zb+12 \zb^2-42 \zb+12 z^2-42 z+24\big) \\
    & \qquad\quad - \log\left((1-z)(1-\zb)\right)\big(4 z^2 \zb^2+14 z^2 \zb+14 z \zb^2+4 z \zb-6 \zb^2-6 z^2\big) \\
    & \qquad\quad + 14 z \zb+2 \zb^2+2 z^2-9 z-9 \zb+6 + \mathcal{O}(\veps) \bigg\}\,,
\end{aligned}
\end{equation}
and, as explained above, this suffices to construct the master integrals $\mi{e}$ and $\mi{j}$.

\section{Permutation group $S_3$}\label{app:permutation-S3}
In Appendix~\ref{app:integrals} we have introduced the MIs associated to the IBPs of the massless three-point, two-loop diagrams.
By looking at the expressions~\eqref{mi1}-\eqref{mi11}, we see that only five integrals are required
\begin{equation}\label{eq:families}
    \mathcal{B}_{1,1}(k)\,,\quad
    \mathcal{B}_{1,1}^{(2)}(k)\,,\quad
    \mathcal{T}(q_1,q_2)\,,\quad
    \left[\mathcal{T}\star\mathcal{B}_{1,1}\right](q_1,q_2)\,,\quad
    \left[\mathcal{T}\star\mathcal{B}_{1,2}\right](q_1,q_2)\,,
\end{equation}
so that all the other integrals are computed from these by exchanges of the external legs.
Here we parametrise the elements of the permutation group of three elements, $S_3$, to make them applicable to the MIs.
The elements of $S_3$ are
\begin{subequations}
\begin{align}
    & g_0 = \mathbb{I}\,,\\
    & g_1 = \{k \to k, q_1\to q_2, q_2\to q_1\} \,, \\
    & g_2 = \{k \to -q_1, q_1\to -k, q_2\to q_2\} \,, \\
    & g_3 = \{k \to -q_2, q_1\to q_1, q_2\to -k\} \,, \\
    & g_4 = \{k \to -q_1, q_1\to q_2, q_2\to -k\} \,, \\
    & g_5 = \{k \to -q_2, q_1\to -k, q_2\to q_1\} \,,
\end{align}
\end{subequations}
in which we recognise $g_{1,2,3}$ to be exchanges and $g_{4,5}$ to be cyclic permutations.
However, we express our MIs in terms of $\left\{ Q,z,\zb \right\}$, requiring to reformulate the group elements as

\begin{subequations}
\begin{align}
    & g_0 = \mathbb{I}\,,\\
    & g_1 = \{Q^2\to Q^2, u\to v, v\to u \} = \{Q^2 \to Q^2, z \to 1-\zb \} \,, \\
    & g_2 = \{Q^2\to uQ^2, u\to \frac{1}{u}, v\to \frac{v}{u} \} = \{Q^2 \to z\zb Q^2, z \to \frac{1}{\zb} \} \,, \\
    & g_3 = \{Q^2\to vQ^2, u\to \frac{u}{v}, v\to \frac{1}{v} \} = \{Q^2 \to (1-z)(1-\zb)Q^2, z \to \frac{\zb}{\zb-1} \} \,, \\
    & g_4 = \{Q^2\to uQ^2, u\to \frac{v}{u}, v\to \frac{1}{u} \} = \{Q^2 \to z\zb Q^2, z\to 1-\frac{1}{z} \} \,, \\
    & g_5 = \{Q^2\to vQ^2, u\to \frac{1}{v}, v\to \frac{u}{v} \} = \{Q^2 \to (1-z)(1-\zb)Q^2, z\to \frac{1}{1-z} \} \,.
\end{align}
\end{subequations}
In particular, using $k=q_1+q_2$ we find the result~\eqref{eq:mi-from-S3}.

\bibliographystyle{JHEP}
\bibliography{references}

@article{LIGOScientific:2026sit,
    author = "Abac, A. G. and others",
    collaboration = "LIGO Scientific, VIRGO, KAGRA",
    title = "{GWTC-5.0: An Introduction to Version 5.0 of the Gravitational-Wave Transient Catalog}",
    eprint = "2605.27223",
    archivePrefix = "arXiv",
    primaryClass = "gr-qc",
    reportNumber = "LIGO-P2500701",
    month = "5",
    year = "2026"
}

@article{ET:2019dnz,
    author = "Maggiore, Michele and others",
    collaboration = "ET",
    title = "{Science Case for the Einstein Telescope}",
    eprint = "1912.02622",
    archivePrefix = "arXiv",
    primaryClass = "astro-ph.CO",
    doi = "10.1088/1475-7516/2020/03/050",
    journal = "JCAP",
    volume = "03",
    pages = "050",
    year = "2020"
}

@article{LISA:2024hlh,
    author = "Colpi, Monica and others",
    collaboration = "LISA",
    title = "{LISA Definition Study Report}",
    eprint = "2402.07571",
    archivePrefix = "arXiv",
    primaryClass = "astro-ph.CO",
    month = "2",
    year = "2024"
}

@article{LISAConsortiumWaveformWorkingGroup:2023arg,
    author = "Afshordi, Niayesh and others",
    collaboration = "LISA Consortium Waveform Working Group",
    title = "{Waveform modelling for the Laser Interferometer Space Antenna}",
    eprint = "2311.01300",
    archivePrefix = "arXiv",
    primaryClass = "gr-qc",
    doi = "10.1007/s41114-025-00056-1",
    journal = "Living Rev. Rel.",
    volume = "28",
    number = "1",
    pages = "9",
    year = "2025"
}

@article{Kovacs:1978eu,
    author = "Kovacs, S. J. and Thorne, K. S.",
    title = "{The Generation of Gravitational Waves. 4. Bremsstrahlung}",
    doi = "10.1086/156350",
    journal = "Astrophys. J.",
    volume = "224",
    pages = "62--85",
    year = "1978"
}

@article{Saavedra:2026ttk,
    author = "Saavedra, Michael",
    title = "{The Classical Gravitational Impulse at High Energies}",
    eprint = "2605.18544",
    archivePrefix = "arXiv",
    primaryClass = "hep-th",
    month = "5",
    year = "2026"
}

@article{Brandhuber:2023hhy,
    author = "Brandhuber, Andreas and Brown, Graham R. and Chen, Gang and De Angelis, Stefano and Gowdy, Joshua and Travaglini, Gabriele",
    title = "{One-loop gravitational bremsstrahlung and waveforms from a heavy-mass effective field theory}",
    eprint = "2303.06111",
    archivePrefix = "arXiv",
    primaryClass = "hep-th",
    reportNumber = "QMUL-22-28,SAGEX-22-32-E",
    doi = "10.1007/JHEP06(2023)048",
    journal = "JHEP",
    volume = "06",
    pages = "048",
    year = "2023"
}

@article{LIGOScientific:2016aoc,
    author = "Abbott, B. P. and others",
    collaboration = "LIGO Scientific, Virgo",
    title = "{Observation of Gravitational Waves from a Binary Black Hole Merger}",
    eprint = "1602.03837",
    archivePrefix = "arXiv",
    primaryClass = "gr-qc",
    reportNumber = "LIGO-P150914",
    doi = "10.1103/PhysRevLett.116.061102",
    journal = "Phys. Rev. Lett.",
    volume = "116",
    number = "6",
    pages = "061102",
    year = "2016"
}

@article{Bern:2019crd,
    author = "Bern, Zvi and Cheung, Clifford and Roiban, Radu and Shen, Chia-Hsien and Solon, Mikhail P. and Zeng, Mao",
    title = "{Black Hole Binary Dynamics from the Double Copy and Effective Theory}",
    eprint = "1908.01493",
    archivePrefix = "arXiv",
    primaryClass = "hep-th",
    reportNumber = "CERN-TH-2019-128, CALT-TH 2019-026, UCLA/TEP/2019/103",
    doi = "10.1007/JHEP10(2019)206",
    journal = "JHEP",
    volume = "10",
    pages = "206",
    year = "2019"
}

@article{Bjerrum-Bohr:2018xdl,
    author = "Bjerrum-Bohr, N. E. J. and Damgaard, Poul H. and Festuccia, Guido and Plant{\'e}, Ludovic and Vanhove, Pierre",
    title = "{General Relativity from Scattering Amplitudes}",
    eprint = "1806.04920",
    archivePrefix = "arXiv",
    primaryClass = "hep-th",
    doi = "10.1103/PhysRevLett.121.171601",
    journal = "Phys. Rev. Lett.",
    volume = "121",
    number = "17",
    pages = "171601",
    year = "2018"
}

@article{DiVecchia:2022piu,
    author = "Di Vecchia, Paolo and Heissenberg, Carlo and Russo, Rodolfo and Veneziano, Gabriele",
    title = "{Classical gravitational observables from the Eikonal operator}",
    eprint = "2210.12118",
    archivePrefix = "arXiv",
    primaryClass = "hep-th",
    reportNumber = "CERN-TH-2022-169, NORDITA 2022-071, QMUL-PH-22-32, UUITP-42/22",
    doi = "10.1016/j.physletb.2023.138049",
    journal = "Phys. Lett. B",
    volume = "843",
    pages = "138049",
    year = "2023"
}

@article{Cheung:2018wkq,
    author = "Cheung, Clifford and Rothstein, Ira Z. and Solon, Mikhail P.",
    title = "{From Scattering Amplitudes to Classical Potentials in the Post-Minkowskian Expansion}",
    eprint = "1808.02489",
    archivePrefix = "arXiv",
    primaryClass = "hep-th",
    reportNumber = "CALT-TH-2018-031",
    doi = "10.1103/PhysRevLett.121.251101",
    journal = "Phys. Rev. Lett.",
    volume = "121",
    number = "25",
    pages = "251101",
    year = "2018"
}

@article{Jakobsen:2022psy,
    author = "Jakobsen, Gustav Uhre and Mogull, Gustav and Plefka, Jan and Sauer, Benjamin",
    title = "{All things retarded: radiation-reaction in worldline quantum field theory}",
    eprint = "2207.00569",
    archivePrefix = "arXiv",
    primaryClass = "hep-th",
    reportNumber = "HU-EP-22/24-RTG",
    doi = "10.1007/JHEP10(2022)128",
    journal = "JHEP",
    volume = "10",
    pages = "128",
    year = "2022"
}

@article{Kalin:2020mvi,
    author = {K{\"a}lin, Gregor and Porto, Rafael A.},
    title = "{Post-Minkowskian Effective Field Theory for Conservative Binary Dynamics}",
    eprint = "2006.01184",
    archivePrefix = "arXiv",
    primaryClass = "hep-th",
    reportNumber = "DESY20-077, SLAC-PUB-17529",
    doi = "10.1007/JHEP11(2020)106",
    journal = "JHEP",
    volume = "11",
    pages = "106",
    year = "2020"
}

@article{Mogull:2020sak,
    author = "Mogull, Gustav and Plefka, Jan and Steinhoff, Jan",
    title = "{Classical black hole scattering from a worldline quantum field theory}",
    eprint = "2010.02865",
    archivePrefix = "arXiv",
    primaryClass = "hep-th",
    reportNumber = "UUITP-37/20, HU-EP-20/22-RTG",
    doi = "10.1007/JHEP02(2021)048",
    journal = "JHEP",
    volume = "02",
    pages = "048",
    year = "2021"
}

@article{Bern:2021dqo,
    author = "Bern, Zvi and Parra-Martinez, Julio and Roiban, Radu and Ruf, Michael S. and Shen, Chia-Hsien and Solon, Mikhail P. and Zeng, Mao",
    title = "{Scattering Amplitudes and Conservative Binary Dynamics at ${\cal O}(G^4)$}",
    eprint = "2101.07254",
    archivePrefix = "arXiv",
    primaryClass = "hep-th",
    reportNumber = "CALT-TH-2021-004, FR-PHENO-2021-03, OUTP-21-03P",
    doi = "10.1103/PhysRevLett.126.171601",
    journal = "Phys. Rev. Lett.",
    volume = "126",
    number = "17",
    pages = "171601",
    year = "2021"
}

@article{Dlapa:2021npj,
    author = {Dlapa, Christoph and K{\"a}lin, Gregor and Liu, Zhengwen and Porto, Rafael A.},
    title = "{Dynamics of binary systems to fourth Post-Minkowskian order from the effective field theory approach}",
    eprint = "2106.08276",
    archivePrefix = "arXiv",
    primaryClass = "hep-th",
    reportNumber = "DESY 21-093, DESY-21-093, MPP-2021-83",
    doi = "10.1016/j.physletb.2022.137203",
    journal = "Phys. Lett. B",
    volume = "831",
    pages = "137203",
    year = "2022"
}

@article{Bern:2021yeh,
    author = "Bern, Zvi and Parra-Martinez, Julio and Roiban, Radu and Ruf, Michael S. and Shen, Chia-Hsien and Solon, Mikhail P. and Zeng, Mao",
    title = "{Scattering Amplitudes, the Tail Effect, and Conservative Binary Dynamics at O(G4)}",
    eprint = "2112.10750",
    archivePrefix = "arXiv",
    primaryClass = "hep-th",
    doi = "10.1103/PhysRevLett.128.161103",
    journal = "Phys. Rev. Lett.",
    volume = "128",
    number = "16",
    pages = "161103",
    year = "2022"
}

@article{Dlapa:2021vgp,
    author = {Dlapa, Christoph and K{\"a}lin, Gregor and Liu, Zhengwen and Porto, Rafael A.},
    title = "{Conservative Dynamics of Binary Systems at Fourth Post-Minkowskian Order in the Large-Eccentricity Expansion}",
    eprint = "2112.11296",
    archivePrefix = "arXiv",
    primaryClass = "hep-th",
    reportNumber = "DESY 21-226",
    doi = "10.1103/PhysRevLett.128.161104",
    journal = "Phys. Rev. Lett.",
    volume = "128",
    number = "16",
    pages = "161104",
    year = "2022"
}

@article{Dlapa:2022lmu,
    author = {Dlapa, Christoph and K{\"a}lin, Gregor and Liu, Zhengwen and Neef, Jakob and Porto, Rafael A.},
    title = "{Radiation Reaction and Gravitational Waves at Fourth Post-Minkowskian Order}",
    eprint = "2210.05541",
    archivePrefix = "arXiv",
    primaryClass = "hep-th",
    doi = "10.1103/PhysRevLett.130.101401",
    journal = "Phys. Rev. Lett.",
    volume = "130",
    number = "10",
    pages = "101401",
    year = "2023"
}

@article{Jakobsen:2023ndj,
    author = "Jakobsen, Gustav Uhre and Mogull, Gustav and Plefka, Jan and Sauer, Benjamin and Xu, Yingxuan",
    title = "{Conservative Scattering of Spinning Black Holes at Fourth Post-Minkowskian Order}",
    eprint = "2306.01714",
    archivePrefix = "arXiv",
    primaryClass = "hep-th",
    reportNumber = "HU-EP-23/16-RTG",
    doi = "10.1103/PhysRevLett.131.151401",
    journal = "Phys. Rev. Lett.",
    volume = "131",
    number = "15",
    pages = "151401",
    year = "2023"
}

@article{Jakobsen:2023hig,
    author = "Jakobsen, Gustav Uhre and Mogull, Gustav and Plefka, Jan and Sauer, Benjamin",
    title = "{Dissipative Scattering of Spinning Black Holes at Fourth Post-Minkowskian Order}",
    eprint = "2308.11514",
    archivePrefix = "arXiv",
    primaryClass = "hep-th",
    reportNumber = "HU-EP-23/47-RTG",
    doi = "10.1103/PhysRevLett.131.241402",
    journal = "Phys. Rev. Lett.",
    volume = "131",
    number = "24",
    pages = "241402",
    year = "2023"
}

@article{Damgaard:2021ipf,
    author = "Damgaard, Poul H. and Plante, Ludovic and Vanhove, Pierre",
    title = "{On an exponential representation of the gravitational S-matrix}",
    eprint = "2107.12891",
    archivePrefix = "arXiv",
    primaryClass = "hep-th",
    reportNumber = "IPhT-t21/037, CERN-TH-2021-111",
    doi = "10.1007/JHEP11(2021)213",
    journal = "JHEP",
    volume = "11",
    pages = "213",
    year = "2021"
}

@article{Dlapa:2024cje,
    author = {Dlapa, Christoph and K{\"a}lin, Gregor and Liu, Zhengwen and Porto, Rafael A.},
    title = "{Local in Time Conservative Binary Dynamics at Fourth Post-Minkowskian Order}",
    eprint = "2403.04853",
    archivePrefix = "arXiv",
    primaryClass = "hep-th",
    reportNumber = "DESY 24-029",
    doi = "10.1103/PhysRevLett.132.221401",
    journal = "Phys. Rev. Lett.",
    volume = "132",
    number = "22",
    pages = "221401",
    year = "2024"
}

@article{Damour:2017zjx,
    author = "Damour, Thibault",
    title = "{High-energy gravitational scattering and the general relativistic two-body problem}",
    eprint = "1710.10599",
    archivePrefix = "arXiv",
    primaryClass = "gr-qc",
    doi = "10.1103/PhysRevD.97.044038",
    journal = "Phys. Rev. D",
    volume = "97",
    number = "4",
    pages = "044038",
    year = "2018"
}

@article{Driesse:2024xad,
    author = "Driesse, Mathias and Jakobsen, Gustav Uhre and Mogull, Gustav and Plefka, Jan and Sauer, Benjamin and Usovitsch, Johann",
    title = "{Conservative Black Hole Scattering at Fifth Post-Minkowskian and First Self-Force Order}",
    eprint = "2403.07781",
    archivePrefix = "arXiv",
    primaryClass = "hep-th",
    reportNumber = "HU-EP-24/08-RTG, CERN-TH-2024-032",
    doi = "10.1103/PhysRevLett.132.241402",
    journal = "Phys. Rev. Lett.",
    volume = "132",
    number = "24",
    pages = "241402",
    year = "2024"
}

@article{Bern:2025wyd,
    author = "Bern, Zvi and Herrmann, Enrico and Roiban, Radu and Ruf, Michael S. and Smirnov, Alexander V. and Smith, Sid and Zeng, Mao",
    title = "{Scattering Amplitudes and Conservative Binary Dynamics at O(G5) without Self-Force Truncation}",
    eprint = "2512.23654",
    archivePrefix = "arXiv",
    primaryClass = "hep-th",
    doi = "10.1103/zjcp-ssrw",
    journal = "Phys. Rev. Lett.",
    volume = "137",
    number = "2",
    pages = "021402",
    year = "2026"
}

@article{Driesse:2026qiz,
    author = "Driesse, Mathias and Jakobsen, Gustav Uhre and Mogull, Gustav and Nega, Christoph and Plefka, Jan and Sauer, Benjamin and Usovitsch, Johann",
    title = "{Conservative Black Hole Scattering at Fifth Post-Minkowskian and Second Self-Force Order}",
    eprint = "2601.16256",
    archivePrefix = "arXiv",
    primaryClass = "hep-th",
    reportNumber = "HU-EP-26/04-RTG",
    doi = "10.1103/4dvk-nglx",
    journal = "Phys. Rev. Lett.",
    volume = "137",
    number = "8",
    pages = "081402",
    year = "2026"
}

@article{Gonzo:2026yha,
    author = "Gonzo, Riccardo and Mogull, Gustav",
    title = "{Canonical Quantisation of Bound and Unbound WQFT}",
    eprint = "2603.05237",
    archivePrefix = "arXiv",
    primaryClass = "hep-th",
    reportNumber = "QMUL-PH-26-06, HU-EP-26/13",
    month = "3",
    year = "2026"
}

@article{Brandhuber:2026dkd,
    author = "Brandhuber, Andreas and Brown, Graham R. and MacDonald, Sebastien and Pichini, Paolo and Travaglini, Gabriele and Vives Matasan, Pablo",
    title = "{The classical limit of the Magnus expansion}",
    eprint = "2609.40291",
    archivePrefix = "arXiv",
    primaryClass = "hep-th",
    reportNumber = "QMUL-PH-26-31",
    month = "9",
    year = "2026"
}

@article{Damgaard:2023ttc,
    author = "Damgaard, Poul H. and Hansen, Elias Roos and Plant{\'e}, Ludovic and Vanhove, Pierre",
    title = "{Classical observables from the exponential representation of the gravitational S-matrix}",
    eprint = "2307.04746",
    archivePrefix = "arXiv",
    primaryClass = "hep-th",
    reportNumber = "CERN-TH-2023-135, IPhT-T23/041, LAPTh-029/23",
    doi = "10.1007/JHEP09(2023)183",
    journal = "JHEP",
    volume = "09",
    pages = "183",
    year = "2023"
}

@article{Brandhuber:2025igz,
    author = "Brandhuber, Andreas and Brown, Graham R. and Pichini, Paolo and Travaglini, Gabriele and Vives Matasan, Pablo",
    title = "{The Magnus expansion in relativistic quantum field theory}",
    eprint = "2512.05017",
    archivePrefix = "arXiv",
    primaryClass = "hep-th",
    reportNumber = "QMUL-PH-25-27",
    doi = "10.1007/JHEP07(2026)151",
    journal = "JHEP",
    volume = "07",
    pages = "151",
    year = "2026"
}

@article{Kovacs:1977uw,
    author = "Kovacs, S. J. and Thorne, K. S.",
    title = "{The Generation of Gravitational Waves. 3. Derivation of Bremsstrahlung Formulas}",
    doi = "10.1086/155576",
    journal = "Astrophys. J.",
    volume = "217",
    pages = "252--280",
    year = "1977"
}

@article{Brunello:2025cot,
    author = "Brunello, Giacomo and Chestnov, Vsevolod and Crisanti, Giulio and Giroux, Mathieu and Smith, Sid",
    title = "{Gravitational waveforms from restriction theory and rapid-decay homology}",
    eprint = "2510.26874",
    archivePrefix = "arXiv",
    primaryClass = "hep-th",
    doi = "10.1103/ynk9-ykw7",
    journal = "Phys. Rev. D",
    volume = "113",
    number = "8",
    pages = "085011",
    year = "2026"
}

@article{DeAngelis:2023lvf,
    author = "De Angelis, Stefano and Novichkov, Pavel P. and Gonzo, Riccardo",
    title = "{Spinning waveforms from the Kosower-Maybee-O{\textquoteright}Connell formalism at leading order}",
    eprint = "2309.17429",
    archivePrefix = "arXiv",
    primaryClass = "hep-th",
    doi = "10.1103/PhysRevD.110.L041502",
    journal = "Phys. Rev. D",
    volume = "110",
    number = "4",
    pages = "L041502",
    year = "2024"
}

@article{Aoude:2023dui,
    author = "Aoude, Rafael and Haddad, Kays and Heissenberg, Carlo and Helset, Andreas",
    title = "{Leading-order gravitational radiation to all spin orders}",
    eprint = "2310.05832",
    archivePrefix = "arXiv",
    primaryClass = "hep-th",
    reportNumber = "CERN-TH-2023-177",
    doi = "10.1103/PhysRevD.109.036007",
    journal = "Phys. Rev. D",
    volume = "109",
    number = "3",
    pages = "036007",
    year = "2024"
}

@article{Brandhuber:2023hhl,
    author = "Brandhuber, Andreas and Brown, Graham R. and Chen, Gang and Gowdy, Joshua and Travaglini, Gabriele",
    title = "{Resummed spinning waveforms from five-point amplitudes}",
    eprint = "2310.04405",
    archivePrefix = "arXiv",
    primaryClass = "hep-th",
    reportNumber = "QMUL-PH-23-18",
    doi = "10.1007/JHEP02(2024)026",
    journal = "JHEP",
    volume = "02",
    pages = "026",
    year = "2024"
}

@article{DiVecchia:2021bdo,
    author = "Di Vecchia, Paolo and Heissenberg, Carlo and Russo, Rodolfo and Veneziano, Gabriele",
    title = "{The eikonal approach to gravitational scattering and radiation at $ \mathcal{O} $(G$^{3}$)}",
    eprint = "2104.03256",
    archivePrefix = "arXiv",
    primaryClass = "hep-th",
    reportNumber = "CERN-TH-2021-046, NORDITA 2021-028, QMUL-PH-21-17",
    doi = "10.1007/JHEP07(2021)169",
    journal = "JHEP",
    volume = "07",
    pages = "169",
    year = "2021"
}

@article{Georgoudis:2023ozp,
    author = "Georgoudis, Alessandro and Heissenberg, Carlo and Vazquez-Holm, Ingrid",
    title = "{Addendum to: Inelastic exponentiation and classical gravitational scattering at one loop}",
    eprint = "2312.14710",
    archivePrefix = "arXiv",
    primaryClass = "hep-th",
    doi = "10.1007/JHEP02(2024)161",
    journal = "JHEP",
    volume = "2024",
    number = "02",
    pages = "161",
    year = "2024"
}

@article{Jakobsen:2021lvp,
    author = "Jakobsen, Gustav Uhre and Mogull, Gustav and Plefka, Jan and Steinhoff, Jan",
    title = "{Gravitational Bremsstrahlung and Hidden Supersymmetry of Spinning Bodies}",
    eprint = "2106.10256",
    archivePrefix = "arXiv",
    primaryClass = "hep-th",
    reportNumber = "HU-EP-21/15-RTG",
    doi = "10.1103/PhysRevLett.128.011101",
    journal = "Phys. Rev. Lett.",
    volume = "128",
    number = "1",
    pages = "011101",
    year = "2022"
}

@article{Bini:2023fiz,
    author = "Bini, Donato and Damour, Thibault and Geralico, Andrea",
    title = "{Comparing one-loop gravitational bremsstrahlung amplitudes to the multipolar-post-Minkowskian waveform}",
    eprint = "2309.14925",
    archivePrefix = "arXiv",
    primaryClass = "gr-qc",
    doi = "10.1103/PhysRevD.108.124052",
    journal = "Phys. Rev. D",
    volume = "108",
    number = "12",
    pages = "124052",
    year = "2023"
}

@article{Bini:2024rsy,
    author = "Bini, Donato and Damour, Thibault and De Angelis, Stefano and Geralico, Andrea and Herderschee, Aidan and Roiban, Radu and Teng, Fei",
    title = "{Gravitational waveforms: A tale of two formalisms}",
    eprint = "2402.06604",
    archivePrefix = "arXiv",
    primaryClass = "hep-th",
    doi = "10.1103/PhysRevD.109.125008",
    journal = "Phys. Rev. D",
    volume = "109",
    number = "12",
    pages = "125008",
    year = "2024"
}

@article{Bohnenblust:2025gir,
    author = "Bohnenblust, Lara and Ita, Harald and Kraus, Manfred and Schlenk, Johannes",
    title = "{Gravitational Bremsstrahlung in Black-Hole Scattering at $\mathcal{O}(G^3)$: Quadratic-in-Spin Effects}",
    eprint = "2505.15724",
    archivePrefix = "arXiv",
    primaryClass = "hep-th",
    doi = "10.1007/JHEP12(2025)100",
    journal = "JHEP",
    volume = "12",
    pages = "100",
    year = "2025"
}

@article{Alessio:2024onn,
    author = "Alessio, Francesco and Di Vecchia, Paolo and Heissenberg, Carlo",
    title = "{Logarithmic soft theorems and soft spectra}",
    eprint = "2407.04128",
    archivePrefix = "arXiv",
    primaryClass = "hep-th",
    doi = "10.1007/JHEP11(2024)124",
    journal = "JHEP",
    volume = "11",
    pages = "124",
    year = "2024"
}

@article{Bini:2026vaq,
    author = "Bini, Donato and Damour, Thibault and De Angelis, Stefano and Geralico, Andrea",
    title = "{Radiated Energy Spectrum, Radiated Angular Distribution and Non-linear Memory from the One-loop Gravitational Bremsstrahlung Waveform}",
    eprint = "2607.03879",
    archivePrefix = "arXiv",
    primaryClass = "gr-qc",
    month = "7",
    year = "2026"
}

@article{Bini:2024ijq,
    author = "Bini, Donato and Damour, Thibault and Geralico, Andrea",
    title = "{Gravitational bremsstrahlung waveform at the fourth post-Minkowskian order and the second post-Newtonian level}",
    eprint = "2407.02076",
    archivePrefix = "arXiv",
    primaryClass = "gr-qc",
    doi = "10.1103/PhysRevD.110.064035",
    journal = "Phys. Rev. D",
    volume = "110",
    number = "6",
    pages = "064035",
    year = "2024"
}

@article{Georgoudis:2025vkk,
    author = "Georgoudis, Alessandro and Goncalves, Vasco and Heissenberg, Carlo and Parra-Martinez, Julio",
    title = "{Nonlinear Gravitational Memory in the Post-Minkowskian Expansion}",
    eprint = "2506.20733",
    archivePrefix = "arXiv",
    primaryClass = "hep-th",
    doi = "10.1103/8m17-s2y8",
    journal = "Phys. Rev. Lett.",
    volume = "136",
    number = "12",
    pages = "121401",
    year = "2026"
}

@article{Bini:2022enm,
    author = "Bini, Donato and Damour, Thibault and Geralico, Andrea",
    title = "{Radiated momentum and radiation reaction in gravitational two-body scattering including time-asymmetric effects}",
    eprint = "2210.07165",
    archivePrefix = "arXiv",
    primaryClass = "gr-qc",
    doi = "10.1103/PhysRevD.107.024012",
    journal = "Phys. Rev. D",
    volume = "107",
    number = "2",
    pages = "024012",
    year = "2023"
}

@article{Swain:2024ngs,
    author = "Swain, Shaun and Pratten, Geraint and Schmidt, Patricia",
    title = "{Strong field scattering of black holes: Assessing resummation strategies}",
    eprint = "2411.09652",
    archivePrefix = "arXiv",
    primaryClass = "gr-qc",
    doi = "10.1103/PhysRevD.111.064048",
    journal = "Phys. Rev. D",
    volume = "111",
    number = "6",
    pages = "064048",
    year = "2025"
}

@article{Abreu:2024xoh,
    author = "Abreu, Samuel and De Laurentis, Giuseppe and Falcioni, Giulio and Gardi, Einan and Milloy, Calum and Vernazza, Leonardo",
    title = "{The two-loop Lipatov vertex in QCD}",
    eprint = "2412.20578",
    archivePrefix = "arXiv",
    primaryClass = "hep-ph",
    reportNumber = "CERN-TH-2024-226, ZU-TH 68/24",
    doi = "10.1007/JHEP04(2025)161",
    journal = "JHEP",
    volume = "04",
    pages = "161",
    year = "2025"
}

@article{Jakobsen:2021smu,
    author = "Jakobsen, Gustav Uhre and Mogull, Gustav and Plefka, Jan and Steinhoff, Jan",
    title = "{Classical Gravitational Bremsstrahlung from a Worldline Quantum Field Theory}",
    eprint = "2101.12688",
    archivePrefix = "arXiv",
    primaryClass = "gr-qc",
    reportNumber = "HU-EP-21/03-RTG",
    doi = "10.1103/PhysRevLett.126.201103",
    journal = "Phys. Rev. Lett.",
    volume = "126",
    number = "20",
    pages = "201103",
    year = "2021"
}

@article{Mougiakakos:2021ckm,
    author = "Mougiakakos, Stavros and Riva, Massimiliano Maria and Vernizzi, Filippo",
    title = "{Gravitational Bremsstrahlung in the post-Minkowskian effective field theory}",
    eprint = "2102.08339",
    archivePrefix = "arXiv",
    primaryClass = "gr-qc",
    doi = "10.1103/PhysRevD.104.024041",
    journal = "Phys. Rev. D",
    volume = "104",
    number = "2",
    pages = "024041",
    year = "2021"
}

@article{Cristofoli:2021vyo,
    author = "Cristofoli, Andrea and Gonzo, Riccardo and Kosower, David A. and O'Connell, Donal",
    title = "{Waveforms from amplitudes}",
    eprint = "2107.10193",
    archivePrefix = "arXiv",
    primaryClass = "hep-th",
    doi = "10.1103/PhysRevD.106.056007",
    journal = "Phys. Rev. D",
    volume = "106",
    number = "5",
    pages = "056007",
    year = "2022"
}

@article{Herderschee:2023fxh,
    author = "Herderschee, Aidan and Roiban, Radu and Teng, Fei",
    title = "{The sub-leading scattering waveform from amplitudes}",
    eprint = "2303.06112",
    archivePrefix = "arXiv",
    primaryClass = "hep-th",
    reportNumber = "LCTP-23-04",
    doi = "10.1007/JHEP06(2023)004",
    journal = "JHEP",
    volume = "06",
    pages = "004",
    year = "2023"
}

@article{Brunello:2025eso,
    author = "Brunello, Giacomo and De Angelis, Stefano and Kosower, David A.",
    title = "{Analytic One-loop Scattering Waveform in General Relativity}",
    eprint = "2511.05412",
    archivePrefix = "arXiv",
    primaryClass = "hep-th",
    doi = "10.1103/6xtl-qn8z",
    journal = "Phys. Rev. D",
    volume = "114",
    number = "4",
    pages = "045004",
    year = "2026"
}

@article{Bohnenblust:2026ujk,
    author = "Bohnenblust, Lara and Jordan Eriksen, Carl and Hoogeveen, Jitze and Jakobsen, Gustav Uhre and Plefka, Jan",
    title = "{Black Hole Response Theory and its Exact Shockwave Limit}",
    eprint = "2604.22009",
    archivePrefix = "arXiv",
    primaryClass = "hep-th",
    reportNumber = "HU-EP-26/16-RTG",
    doi = "10.1007/JHEP08(2026)064",
    journal = "JHEP",
    volume = "08",
    pages = "064",
    year = "2026"
}

@article{Georgoudis:2023eke,
    author = "Georgoudis, Alessandro and Heissenberg, Carlo and Russo, Rodolfo",
    title = "{An eikonal-inspired approach to the gravitational scattering waveform}",
    eprint = "2312.07452",
    archivePrefix = "arXiv",
    primaryClass = "hep-th",
    reportNumber = "QMUL-PH-23-34",
    doi = "10.1007/JHEP03(2024)089",
    journal = "JHEP",
    volume = "03",
    pages = "089",
    year = "2024"
}

@article{Kotikov:2018wxe,
    author = "Kotikov, A. V. and Teber, S.",
    title = "{Multi-loop techniques for massless Feynman diagram calculations}",
    eprint = "1805.05109",
    archivePrefix = "arXiv",
    primaryClass = "hep-th",
    doi = "10.1134/S1063779619010039",
    journal = "Phys. Part. Nucl.",
    volume = "50",
    number = "1",
    pages = "1--41",
    year = "2019"
}

@article{Bern:2020gjj,
    author = "Bern, Zvi and Ita, Harald and Parra-Martinez, Julio and Ruf, Michael S.",
    title = "{Universality in the classical limit of massless gravitational scattering}",
    eprint = "2002.02459",
    archivePrefix = "arXiv",
    primaryClass = "hep-th",
    reportNumber = "UCLA/TEP/2020/101, FR-PHENO-2020-001",
    doi = "10.1103/PhysRevLett.125.031601",
    journal = "Phys. Rev. Lett.",
    volume = "125",
    number = "3",
    pages = "031601",
    year = "2020"
}

@article{Caron-Huot:2023vxl,
    author = "Caron-Huot, Simon and Giroux, Mathieu and Hannesdottir, Holmfridur S. and Mizera, Sebastian",
    title = "{What can be measured asymptotically?}",
    eprint = "2308.02125",
    archivePrefix = "arXiv",
    primaryClass = "hep-th",
    doi = "10.1007/JHEP01(2024)139",
    journal = "JHEP",
    volume = "01",
    pages = "139",
    year = "2024"
}

@article{Bohnenblust:2023qmy,
    author = "Bohnenblust, Lara and Ita, Harald and Kraus, Manfred and Schlenk, Johannes",
    title = "{Gravitational Bremsstrahlung in black-hole scattering at $ \mathcal{O}\left({G}^3\right) $: linear-in-spin effects}",
    eprint = "2312.14859",
    archivePrefix = "arXiv",
    primaryClass = "hep-th",
    reportNumber = "PSI-PR-23-47",
    doi = "10.1007/JHEP11(2024)109",
    journal = "JHEP",
    volume = "11",
    pages = "109",
    year = "2024"
}

@article{Elkhidir:2023dco,
    author = "Elkhidir, Asaad and O'Connell, Donal and Sergola, Matteo and Vazquez-Holm, Ingrid A.",
    title = "{Radiation and reaction at one loop}",
    eprint = "2303.06211",
    archivePrefix = "arXiv",
    primaryClass = "hep-th",
    doi = "10.1007/JHEP07(2024)272",
    journal = "JHEP",
    volume = "07",
    pages = "272",
    year = "2024"
}

@article{Amati:2007ak,
    author = "Amati, D. and Ciafaloni, M. and Veneziano, G.",
    title = "{Towards an S-matrix description of gravitational collapse}",
    eprint = "0712.1209",
    archivePrefix = "arXiv",
    primaryClass = "hep-th",
    reportNumber = "CERN-PH-TH-2007-254",
    doi = "10.1088/1126-6708/2008/02/049",
    journal = "JHEP",
    volume = "02",
    pages = "049",
    year = "2008"
}

@article{Amati:1993tb,
    author = "Amati, D. and Ciafaloni, M. and Veneziano, G.",
    title = "{Effective action and all order gravitational eikonal at Planckian energies}",
    reportNumber = "CERN-TH-6791-93, SISSA-11-93-EP",
    doi = "10.1016/0550-3213(93)90367-X",
    journal = "Nucl. Phys. B",
    volume = "403",
    pages = "707--724",
    year = "1993"
}

@article{Georgoudis:2023lgf,
    author = "Georgoudis, Alessandro and Heissenberg, Carlo and Vazquez-Holm, Ingrid",
    title = "{Inelastic exponentiation and classical gravitational scattering at one loop}",
    eprint = "2303.07006",
    archivePrefix = "arXiv",
    primaryClass = "hep-th",
    reportNumber = "NORDITA 2023-010, UUITP-03/23",
    doi = "10.1007/JHEP06(2023)126",
    journal = "JHEP",
    volume = "2023",
    number = "06",
    pages = "126",
    year = "2023"
}

@article{Alessio:2026bdi,
    author = "Alessio, Francesco and Del Duca, Vittorio and Gonzo, Riccardo and Rosi, Emanuele",
    title = "{Gravitational amplitudes in the Regge limit: waveforms, shock waves and unitarity cuts}",
    eprint = "2601.21687",
    archivePrefix = "arXiv",
    primaryClass = "hep-th",
    reportNumber = "QMUL-PH-26-03",
    month = "1",
    year = "2026"
}

@article{Amati:1990xe,
    author = "Amati, D. and Ciafaloni, M. and Veneziano, G.",
    title = "{Higher Order Gravitational Deflection and Soft Bremsstrahlung in Planckian Energy Superstring Collisions}",
    reportNumber = "CERN-TH-5636-90",
    doi = "10.1016/0550-3213(90)90375-N",
    journal = "Nucl. Phys. B",
    volume = "347",
    pages = "550--580",
    year = "1990"
}

@article{Alessio:2025isu,
    author = "Alessio, Francesco and Del Duca, Vittorio and Gonzo, Riccardo and Rosi, Emanuele and Rothstein, Ira Z. and Saavedra, Michael",
    title = "{Analytic structure of the high-energy gravitational amplitude: Multi-H diagrams and classical 5PM logarithms}",
    eprint = "2511.11457",
    archivePrefix = "arXiv",
    primaryClass = "hep-th",
    reportNumber = "QMUL-PH-25-21",
    doi = "10.1103/2qtg-6msp",
    journal = "Phys. Rev. D",
    volume = "113",
    number = "10",
    pages = "105013",
    year = "2026"
}

@article{Balitsky:1978ic,
    author = "Balitsky, I. I. and Lipatov, L. N.",
    title = "{The Pomeranchuk Singularity in Quantum Chromodynamics}",
    journal = "Sov. J. Nucl. Phys.",
    volume = "28",
    pages = "822--829",
    year = "1978"
}

@article{Balitsky:1995ub,
    author = "Balitsky, I.",
    title = "{Operator expansion for high-energy scattering}",
    eprint = "hep-ph/9509348",
    archivePrefix = "arXiv",
    reportNumber = "MIT-CTP-2470",
    doi = "10.1016/0550-3213(95)00638-9",
    journal = "Nucl. Phys. B",
    volume = "463",
    pages = "99--160",
    year = "1996"
}

@article{Jalilian-Marian:1997jhx,
    author = "Jalilian-Marian, Jamal and Kovner, Alex and Leonidov, Andrei and Weigert, Heribert",
    title = "{The Wilson renormalization group for low x physics: Towards the high density regime}",
    eprint = "hep-ph/9706377",
    archivePrefix = "arXiv",
    reportNumber = "TPI-MINN-97-20-T, NUC-MINN-97-6-T, HEP-MINN-97-1546, CAVENDISH-HEP-97-09",
    doi = "10.1103/PhysRevD.59.014014",
    journal = "Phys. Rev. D",
    volume = "59",
    pages = "014014",
    year = "1998"
}

@article{Jalilian-Marian:1996mkd,
    author = "Jalilian-Marian, Jamal and Kovner, Alex and McLerran, Larry D. and Weigert, Heribert",
    title = "{The Intrinsic glue distribution at very small x}",
    eprint = "hep-ph/9606337",
    archivePrefix = "arXiv",
    reportNumber = "TPI-MINN-96-1429, NUC-MINN-96-10-T, TPI-MINN-96-08",
    doi = "10.1103/PhysRevD.55.5414",
    journal = "Phys. Rev. D",
    volume = "55",
    pages = "5414--5428",
    year = "1997"
}

@article{Kovchegov:1999yj,
    author = "Kovchegov, Yuri V.",
    title = "{Small x F(2) structure function of a nucleus including multiple pomeron exchanges}",
    eprint = "hep-ph/9901281",
    archivePrefix = "arXiv",
    reportNumber = "NUC-MN-99-1-T, TPI-MINN-99-05",
    doi = "10.1103/PhysRevD.60.034008",
    journal = "Phys. Rev. D",
    volume = "60",
    pages = "034008",
    year = "1999"
}

@article{Mueller:1994jq,
    author = "Mueller, Alfred H. and Patel, Bimal",
    title = "{Single and double BFKL pomeron exchange and a dipole picture of high-energy hard processes}",
    eprint = "hep-ph/9403256",
    archivePrefix = "arXiv",
    reportNumber = "CU-TP-625",
    doi = "10.1016/0550-3213(94)90284-4",
    journal = "Nucl. Phys. B",
    volume = "425",
    pages = "471--488",
    year = "1994"
}

@article{Beneke:2012xa,
    author = "Beneke, Martin and Kirilin, Grisha",
    title = "{Soft-collinear gravity}",
    eprint = "1207.4926",
    archivePrefix = "arXiv",
    primaryClass = "hep-ph",
    reportNumber = "TUM-HEP-848-12, TTK-12-31",
    doi = "10.1007/JHEP09(2012)066",
    journal = "JHEP",
    volume = "09",
    pages = "066",
    year = "2012"
}

@article{Bern:2026oqp,
    author = "Bern, Zvi and Jackman, Avery and Mansfield, Gareth and Ruf, Michael S.",
    title = "{Classical Gravitational Scattering from the Ultraviolet and the Absence of Calabi-Yau Integrals in the Conservative Sector at $O(G^5)$}",
    eprint = "2603.15383",
    archivePrefix = "arXiv",
    primaryClass = "hep-th",
    month = "3",
    year = "2026"
}

@article{Kalin:2019rwq,
    author = {K{\"a}lin, Gregor and Porto, Rafael A.},
    title = "{From Boundary Data to Bound States}",
    eprint = "1910.03008",
    archivePrefix = "arXiv",
    primaryClass = "hep-th",
    reportNumber = "DESY 19-167, UUITP-40/19, DESY-19-167",
    doi = "10.1007/JHEP01(2020)072",
    journal = "JHEP",
    volume = "01",
    pages = "072",
    year = "2020"
}

@article{Kalin:2019inp,
    author = {K{\"a}lin, Gregor and Porto, Rafael A.},
    title = "{From boundary data to bound states. Part II. Scattering angle to dynamical invariants (with twist)}",
    eprint = "1911.09130",
    archivePrefix = "arXiv",
    primaryClass = "hep-th",
    reportNumber = "DESY 19-201, SLAC-PUB-17487, UUITP-46/19, DESY-19-201",
    doi = "10.1007/JHEP02(2020)120",
    journal = "JHEP",
    volume = "02",
    pages = "120",
    year = "2020"
}

@article{Cho:2021arx,
    author = {Cho, Gihyuk and K{\"a}lin, Gregor and Porto, Rafael A.},
    title = "{From boundary data to bound states. Part III. Radiative effects}",
    eprint = "2112.03976",
    archivePrefix = "arXiv",
    primaryClass = "hep-th",
    reportNumber = "DESY 21-212",
    doi = "10.1007/JHEP04(2022)154",
    journal = "JHEP",
    volume = "04",
    pages = "154",
    year = "2022",
    note = "[Erratum: JHEP 07, 002 (2022)]"
}

@article{Iancu:2001ad,
    author = "Iancu, Edmond and Leonidov, Andrei and McLerran, Larry D.",
    title = "{The Renormalization group equation for the color glass condensate}",
    eprint = "hep-ph/0102009",
    archivePrefix = "arXiv",
    reportNumber = "BNL-NT-01-3",
    doi = "10.1016/S0370-2693(01)00524-X",
    journal = "Phys. Lett. B",
    volume = "510",
    pages = "133--144",
    year = "2001"
}

@article{Buccioni:2024gzo,
    author = "Buccioni, Federico and Caola, Fabrizio and Devoto, Federica and Gambuti, Giulio",
    title = "{Investigating the universality of five-point QCD scattering amplitudes at high energy}",
    eprint = "2411.14050",
    archivePrefix = "arXiv",
    primaryClass = "hep-ph",
    reportNumber = "OUTP-24-06P, SLAC-PUB-241120, TUM-HEP-1537/24",
    doi = "10.1007/JHEP03(2025)129",
    journal = "JHEP",
    volume = "03",
    pages = "129",
    year = "2025"
}

@article{Caron-Huot:2020vlo,
    author = "Caron-Huot, Simon and Chicherin, Dmitry and Henn, Johannes and Zhang, Yang and Zoia, Simone",
    title = "{Multi-Regge Limit of the Two-Loop Five-Point Amplitudes in $\mathcal{N} = 4$ Super Yang-Mills and $\mathcal{N} = 8$ Supergravity}",
    eprint = "2003.03120",
    archivePrefix = "arXiv",
    primaryClass = "hep-th",
    doi = "10.1007/JHEP10(2020)188",
    journal = "JHEP",
    volume = "10",
    pages = "188",
    year = "2020"
}

@article{Caron-Huot:2013fea,
    author = "Caron-Huot, Simon",
    title = "{When does the gluon reggeize?}",
    eprint = "1309.6521",
    archivePrefix = "arXiv",
    primaryClass = "hep-th",
    doi = "10.1007/JHEP05(2015)093",
    journal = "JHEP",
    volume = "05",
    pages = "093",
    year = "2015"
}

@article{Caron-Huot:2017fxr,
    author = "Caron-Huot, Simon and Gardi, Einan and Vernazza, Leonardo",
    title = "{Two-parton scattering in the high-energy limit}",
    eprint = "1701.05241",
    archivePrefix = "arXiv",
    primaryClass = "hep-ph",
    reportNumber = "EDINBURGH-2017-01",
    doi = "10.1007/JHEP06(2017)016",
    journal = "JHEP",
    volume = "06",
    pages = "016",
    year = "2017"
}

@article{Blackstad:2026eoa,
    author = "Blackstad, Isabelle and Raj, Himanshu and Venugopalan, Raju",
    title = "{Tidal effects on radiation in gravitational shockwave collisions}",
    eprint = "2608.27585",
    archivePrefix = "arXiv",
    primaryClass = "hep-th",
    month = "8",
    year = "2026"
}

@article{Derkachov:1990osa,
    author = "Derkachov, S. E. and Honkonen, J. and Pis'mak, Y. M.",
    title = "{Three-loop calculation of the random walk problem: an application of dimensional transformation and the uniqueness method}",
    doi = "10.1088/0305-4470/23/23/028",
    journal = "J. Phys. A:Math. Gen.",
    volume = "23",
    number = "23",
    pages = "5563--5576",
    year = "1990"
}

@article{Tarasov:1996br,
    author = "Tarasov, O. V.",
    title = "{Connection between Feynman integrals having different values of the space-time dimension}",
    eprint = "hep-th/9606018",
    archivePrefix = "arXiv",
    reportNumber = "DESY-96-068, JINR-E2-96-62",
    doi = "10.1103/PhysRevD.54.6479",
    journal = "Phys. Rev. D",
    volume = "54",
    pages = "6479--6490",
    year = "1996"
}

@article{Lee:2010wea,
    author = "Lee, R. N.",
    editor = {Bl{\"u}mlein, Johannes and Moch, Sven-Olaf and Riemann, Tord},
    title = "{Calculating multiloop integrals using dimensional recurrence relation and $D$-analyticity}",
    eprint = "1007.2256",
    archivePrefix = "arXiv",
    primaryClass = "hep-ph",
    doi = "10.1016/j.nuclphysbps.2010.08.032",
    journal = "Nucl. Phys. B Proc. Suppl.",
    volume = "205-206",
    pages = "135--140",
    year = "2010"
}

@article{Duhr:2019tlz,
    author = "Duhr, Claude and Dulat, Falko",
    title = "{PolyLogTools {\textemdash} polylogs for the masses}",
    eprint = "1904.07279",
    archivePrefix = "arXiv",
    primaryClass = "hep-th",
    reportNumber = "CP3-19-17, CERN-TH-2019-045, SLAC-PUB-17423",
    doi = "10.1007/JHEP08(2019)135",
    journal = "JHEP",
    volume = "08",
    pages = "135",
    year = "2019"
}

@article{Chavez:2012kn,
    author = "Chavez, Federico and Duhr, Claude",
    title = "{Three-mass triangle integrals and single-valued polylogarithms}",
    eprint = "1209.2722",
    archivePrefix = "arXiv",
    primaryClass = "hep-ph",
    doi = "10.1007/JHEP11(2012)114",
    journal = "JHEP",
    volume = "11",
    pages = "114",
    year = "2012"
}

@article{Chicherin:2020oor,
    author = "Chicherin, Dmitry and Sotnikov, Vasily",
    title = "{Pentagon Functions for Scattering of Five Massless Particles}",
    eprint = "2009.07803",
    archivePrefix = "arXiv",
    primaryClass = "hep-ph",
    reportNumber = "MPP-2020-171",
    doi = "10.1007/JHEP12(2020)167",
    journal = "JHEP",
    volume = "20",
    pages = "167",
    year = "2020"
}

@article{DelDuca:1995hf,
    author = "Del Duca, Vittorio",
    title = "{An introduction to the perturbative QCD pomeron and to jet physics at large rapidities}",
    eprint = "hep-ph/9503226",
    archivePrefix = "arXiv",
    reportNumber = "DESY-95-023, DFTT-13-95",
    month = "2",
    year = "1995"
}

@article{DelDuca:2019tur,
    author = "Del Duca, V. and Druc, S. and Drummond, J. M. and Duhr, C. and Dulat, F. and Marzucca, R. and Papathanasiou, G. and Verbeek, B.",
    title = "{All-order amplitudes at any multiplicity in the multi-Regge limit}",
    eprint = "1912.00188",
    archivePrefix = "arXiv",
    primaryClass = "hep-th",
    reportNumber = "DESY-19-180, CERN-TH-2019-204, IPPP/19/86, UUITP-48/19, SLAC-PUB-17491",
    doi = "10.1103/PhysRevLett.124.161602",
    journal = "Phys. Rev. Lett.",
    volume = "124",
    number = "16",
    pages = "161602",
    year = "2020"
}

@article{DiVecchia:2020ymx,
    author = "Di Vecchia, Paolo and Heissenberg, Carlo and Russo, Rodolfo and Veneziano, Gabriele",
    title = "{Universality of ultra-relativistic gravitational scattering}",
    eprint = "2008.12743",
    archivePrefix = "arXiv",
    primaryClass = "hep-th",
    reportNumber = "CERN-TH-2020-143, NORDITA 2020-079, QMUL-PH-20-25",
    doi = "10.1016/j.physletb.2020.135924",
    journal = "Phys. Lett. B",
    volume = "811",
    pages = "135924",
    year = "2020"
}

@article{DiVecchia:2023frv,
    author = "Di Vecchia, Paolo and Heissenberg, Carlo and Russo, Rodolfo and Veneziano, Gabriele",
    title = "{The gravitational eikonal: From particle, string and brane collisions to black-hole encounters}",
    eprint = "2306.16488",
    archivePrefix = "arXiv",
    primaryClass = "hep-th",
    reportNumber = "CERN-TH-2023-108, NORDITA 2023-026, QMUL-PH-23-09, UUITP-14/23",
    doi = "10.1016/j.physrep.2024.06.002",
    journal = "Phys. Rept.",
    volume = "1083",
    pages = "1--169",
    year = "2024"
}

@article{Long:2024ltn,
    author = "Long, Oliver and Whittall, Christopher and Barack, Leor",
    title = "{Black hole scattering near the transition to plunge: Self-force and resummation of post-Minkowskian theory}",
    eprint = "2406.08363",
    archivePrefix = "arXiv",
    primaryClass = "gr-qc",
    doi = "10.1103/PhysRevD.110.044039",
    journal = "Phys. Rev. D",
    volume = "110",
    number = "4",
    pages = "044039",
    year = "2024"
}

@article{Damour:2022ybd,
    author = "Damour, Thibault and Rettegno, Piero",
    title = "{Strong-field scattering of two black holes: Numerical relativity meets post-Minkowskian gravity}",
    eprint = "2211.01399",
    archivePrefix = "arXiv",
    primaryClass = "gr-qc",
    doi = "10.1103/PhysRevD.107.064051",
    journal = "Phys. Rev. D",
    volume = "107",
    number = "6",
    pages = "064051",
    year = "2023"
}

@article{Long:2025nmj,
    author = {Long, Oliver and Pfeiffer, Harald P. and Buonanno, Alessandra and Jakobsen, Gustav Uhre and Mogull, Gustav and Ramos-Buades, Antoni and R{\"u}ter, Hannes R. and Kidder, Lawrence E. and Scheel, Mark A.},
    title = "{Highly accurate simulations of asymmetric black-hole scattering and cross validation of effective-one-body models}",
    eprint = "2507.08071",
    archivePrefix = "arXiv",
    primaryClass = "gr-qc",
    doi = "10.1103/p6fx-7798",
    journal = "Phys. Rev. D",
    volume = "112",
    number = "12",
    pages = "124039",
    year = "2025"
}

@article{Falcioni:2021buo,
    author = "Falcioni, Giulio and Gardi, Einan and Maher, Niamh and Milloy, Calum and Vernazza, Leonardo",
    title = "{Scattering amplitudes in the Regge limit and the soft anomalous dimension through four loops}",
    eprint = "2111.10664",
    archivePrefix = "arXiv",
    primaryClass = "hep-ph",
    reportNumber = "CERN-TH-2021-200",
    doi = "10.1007/JHEP03(2022)053",
    journal = "JHEP",
    volume = "03",
    pages = "053",
    year = "2022"
}

@book{Gribov:2003nw,
    author = "Gribov, V. N.",
    title = "{The theory of complex angular momenta: Gribov lectures on theoretical physics}",
    doi = "10.1017/CBO9780511534959",
    isbn = "978-0-521-03703-7, 978-0-521-81834-6, 978-0-511-05504-1",
    publisher = "Cambridge University Press",
    series = "Cambridge Monographs on Mathematical Physics",
    month = "6",
    year = "2007"
}

@article{Kosower:2018adc,
    author = "Kosower, David A. and Maybee, Ben and O'Connell, Donal",
    title = "{Amplitudes, Observables, and Classical Scattering}",
    eprint = "1811.10950",
    archivePrefix = "arXiv",
    primaryClass = "hep-th",
    doi = "10.1007/JHEP02(2019)137",
    journal = "JHEP",
    volume = "02",
    pages = "137",
    year = "2019"
}

@article{Kuraev:1976ge,
    author = "Kuraev, E. A. and Lipatov, L. N. and Fadin, Victor S.",
    title = "{Multi - Reggeon Processes in the Yang-Mills Theory}",
    journal = "Sov. Phys. JETP",
    volume = "44",
    pages = "443--450",
    year = "1976"
}

@article{Kuraev:1977fs,
    author = "Kuraev, E. A. and Lipatov, L. N. and Fadin, Victor S.",
    title = "{The Pomeranchuk Singularity in Nonabelian Gauge Theories}",
    journal = "Sov. Phys. JETP",
    volume = "45",
    pages = "199--204",
    year = "1977"
}

@article{Lee:2013mka,
    author = "Lee, Roman N.",
    editor = "Wang, Jianxiong",
    title = "{LiteRed 1.4: a powerful tool for reduction of multiloop integrals}",
    eprint = "1310.1145",
    archivePrefix = "arXiv",
    primaryClass = "hep-ph",
    doi = "10.1088/1742-6596/523/1/012059",
    journal = "J. Phys. Conf. Ser.",
    volume = "523",
    pages = "012059",
    year = "2014"
}

@article{Lipatov:1976zz,
    author = "Lipatov, L. N.",
    title = "{Reggeization of the Vector Meson and the Vacuum Singularity in Nonabelian Gauge Theories}",
    journal = "Sov. J. Nucl. Phys.",
    volume = "23",
    pages = "338--345",
    year = "1976"
}

@article{Lipatov:1982it,
    author = "Lipatov, L. N.",
    title = "{Multi - Regge Processes in Gravitation}",
    reportNumber = "LENINGRAD-82-747",
    journal = "Sov. Phys. JETP",
    volume = "55",
    pages = "582--590",
    year = "1982"
}

@article{Lipatov:1989bs,
    author = "Lipatov, L. N.",
    title = "{POMERON IN QUANTUM CHROMODYNAMICS}",
    doi = "10.1142/9789814503266_0004",
    journal = "Adv. Ser. Direct. High Energy Phys.",
    volume = "5",
    pages = "411--489",
    year = "1989"
}

@article{Lipatov:1993qn,
    author = "Lipatov, L. N.",
    editor = "Tran Thanh Van, J.",
    title = "{High-energy asymptotics of multicolor QCD and two-dimensional conformal field theories}",
    reportNumber = "DESY-93-055",
    doi = "10.1016/0370-2693(93)90951-D",
    journal = "Phys. Lett. B",
    volume = "309",
    pages = "394--396",
    year = "1993"
}

@article{Lipatov:1993yb,
    author = "Lipatov, L. N.",
    title = "{Asymptotic behavior of multicolor QCD at high energies in connection with exactly solvable spin models}",
    eprint = "hep-th/9311037",
    archivePrefix = "arXiv",
    reportNumber = "DFPD-93-TH-70",
    journal = "JETP Lett.",
    volume = "59",
    pages = "596--599",
    year = "1994"
}

@article{Lipatov:2009nt,
    author = "Lipatov, L. N.",
    title = "{Integrability of scattering amplitudes in N=4 SUSY}",
    eprint = "0902.1444",
    archivePrefix = "arXiv",
    primaryClass = "hep-th",
    reportNumber = "DESY-09-020",
    doi = "10.1088/1751-8113/42/30/304020",
    journal = "J. Phys. A",
    volume = "42",
    pages = "304020",
    year = "2009"
}

@article{White:2011yy,
    author = "White, Chris D.",
    title = "{Factorization Properties of Soft Graviton Amplitudes}",
    eprint = "1103.2981",
    archivePrefix = "arXiv",
    primaryClass = "hep-th",
    doi = "10.1007/JHEP05(2011)060",
    journal = "JHEP",
    volume = "05",
    pages = "060",
    year = "2011"
}

@article{Melville:2013qca,
    author = "Melville, S. and Naculich, S. G. and Schnitzer, H. J. and White, C. D.",
    title = "{Wilson line approach to gravity in the high energy limit}",
    eprint = "1306.6019",
    archivePrefix = "arXiv",
    primaryClass = "hep-th",
    reportNumber = "BRX-TH-667, BOW-PH-157",
    doi = "10.1103/PhysRevD.89.025009",
    journal = "Phys. Rev. D",
    volume = "89",
    number = "2",
    pages = "025009",
    year = "2014"
}

@article{Okui:2017all,
    author = "Okui, Takemichi and Yunesi, Arash",
    title = "{Soft collinear effective theory for gravity}",
    eprint = "1710.07685",
    archivePrefix = "arXiv",
    primaryClass = "hep-th",
    doi = "10.1103/PhysRevD.97.066011",
    journal = "Phys. Rev. D",
    volume = "97",
    number = "6",
    pages = "066011",
    year = "2018"
}

@article{Raj:2024xsi,
    author = "Raj, Himanshu and Venugopalan, Raju",
    title = "{QCD-gravity double-copy in the Regge regime: Shock wave propagators}",
    eprint = "2406.10483",
    archivePrefix = "arXiv",
    primaryClass = "hep-th",
    doi = "10.1103/PhysRevD.110.056010",
    journal = "Phys. Rev. D",
    volume = "110",
    number = "5",
    pages = "056010",
    year = "2024"
}

@article{Usyukina:1994iw,
    author = "Usyukina, Natalia I. and Davydychev, Andrei I.",
    title = "{New results for two loop off-shell three point diagrams}",
    eprint = "hep-ph/9402223",
    archivePrefix = "arXiv",
    reportNumber = "BERGEN-1994-03",
    doi = "10.1016/0370-2693(94)90874-5",
    journal = "Phys. Lett. B",
    volume = "332",
    pages = "159--167",
    year = "1994"
}

@article{Birthwright:2004kk,
    author = "Birthwright, T. G. and Glover, E. W. Nigel and Marquard, P.",
    title = "{Master integrals for massless two-loop vertex diagrams with three offshell legs}",
    eprint = "hep-ph/0407343",
    archivePrefix = "arXiv",
    reportNumber = "DCPT-04-90, IPPP-04-45",
    doi = "10.1088/1126-6708/2004/09/042",
    journal = "JHEP",
    volume = "09",
    pages = "042",
    year = "2004"
}

@article{Rothstein:2024nlq,
    author = "Rothstein, Ira Z. and Saavedra, Michael",
    title = "{A Systematic Lagrangian Formulation for Quantum and Classical Gravity at High Energies}",
    eprint = "2412.04428",
    archivePrefix = "arXiv",
    primaryClass = "hep-th",
    month = "12",
    year = "2024"
}

@article{Weinberg:1964ew,
    author = "Weinberg, Steven",
    title = "{Photons and Gravitons in  $S$-Matrix Theory: Derivation of Charge Conservation and Equality of Gravitational and Inertial Mass}",
    doi = "10.1103/PhysRev.135.B1049",
    journal = "Phys. Rev.",
    volume = "135",
    pages = "B1049--B1056",
    year = "1964"
}

@article{Chiu:2012ir,
    author = "Chiu, Jui-Yu and Jain, Ambar and Neill, Duff and Rothstein, Ira Z.",
    title = "{A Formalism for the Systematic Treatment of Rapidity Logarithms in Quantum Field Theory}",
    eprint = "1202.0814",
    archivePrefix = "arXiv",
    primaryClass = "hep-ph",
    reportNumber = "INT-PUB-11-057, CALT-68-2864",
    doi = "10.1007/JHEP05(2012)084",
    journal = "JHEP",
    volume = "05",
    pages = "084",
    year = "2012"
}

@article{Naculich:2011ry,
    author = "Naculich, Stephen G. and Schnitzer, Howard J.",
    title = "{Eikonal methods applied to gravitational scattering amplitudes}",
    eprint = "1101.1524",
    archivePrefix = "arXiv",
    primaryClass = "hep-th",
    reportNumber = "BOW-PH-148, BRX-TH-628",
    doi = "10.1007/JHEP05(2011)087",
    journal = "JHEP",
    volume = "05",
    pages = "087",
    year = "2011"
}

@article{Ciafaloni:2015xsr,
    author = "Ciafaloni, Marcello and Colferai, Dimitri and Coradeschi, Francesco and Veneziano, Gabriele",
    title = "{Unified limiting form of graviton radiation at extreme energies}",
    eprint = "1512.00281",
    archivePrefix = "arXiv",
    primaryClass = "hep-th",
    reportNumber = "CERN-PH-TH-2015-272",
    doi = "10.1103/PhysRevD.93.044052",
    journal = "Phys. Rev. D",
    volume = "93",
    number = "4",
    pages = "044052",
    year = "2016"
}

@article{Weinberg:1965nx,
    author = "Weinberg, Steven",
    title = "{Infrared photons and gravitons}",
    doi = "10.1103/PhysRev.140.B516",
    journal = "Phys. Rev.",
    volume = "140",
    pages = "B516--B524",
    year = "1965"
}

@book{Weinzierl:2022eaz,
    author = "Weinzierl, Stefan",
    title = "{Feynman Integrals. A Comprehensive Treatment for Students and Researchers}",
    eprint = "2201.03593",
    archivePrefix = "arXiv",
    primaryClass = "hep-th",
    reportNumber = "MITP/22-001",
    doi = "10.1007/978-3-030-99558-4",
    isbn = "978-3-030-99557-7, 978-3-030-99560-7, 978-3-030-99558-4",
    publisher = "Springer",
    series = "UNITEXT for Physics",
    year = "2022"
}

@article{Rosi:2026zjp,
    author = "Rosi, Emanuele",
    title = "{Gravitational Radiation from Shockwave Scattering in the Self-Force EFT}",
    eprint = "2609.09378",
    archivePrefix = "arXiv",
    primaryClass = "gr-qc",
    month = "9",
    year = "2026"
}

@article{Raj:2025hse,
    author = "Raj, H. and Venugopalan, R.",
    title = "{QCD-Gravity double copy in Regge asymptotics: from $2\rightarrow n$ amplitudes to radiation in shockwave collisions}",
    eprint = "2507.21252",
    archivePrefix = "arXiv",
    primaryClass = "hep-th",
    doi = "10.5506/APhysPolB.56.11-A1",
    journal = "Acta Phys. Polon. B",
    volume = "56",
    number = "11",
    pages = "11--A1",
    year = "2025"
}

@article{Raj:2023irr,
    author = "Raj, Himanshu and Venugopalan, Raju",
    title = "{Universal features of 2{\textrightarrow}N scattering in QCD and gravity from shockwave collisions}",
    eprint = "2311.03463",
    archivePrefix = "arXiv",
    primaryClass = "hep-th",
    doi = "10.1103/PhysRevD.109.044064",
    journal = "Phys. Rev. D",
    volume = "109",
    number = "4",
    pages = "044064",
    year = "2024"
}

@article{Gehrmann:2018yef,
    author = "Gehrmann, T. and Henn, J. M. and Lo Presti, N. A.",
    title = "{Pentagon functions for massless planar scattering amplitudes}",
    eprint = "1807.09812",
    archivePrefix = "arXiv",
    primaryClass = "hep-ph",
    reportNumber = "ZU-TH-25-18, ZU-TH 25/18, MPP-2018-180, MITP-18-067, IPPP-18-59, MITP/18-067, IPPP/18/59",
    doi = "10.1007/JHEP10(2018)103",
    journal = "JHEP",
    volume = "10",
    pages = "103",
    year = "2018"
}

@article{Alessio:2025flu,
    author = "Alessio, Francesco and Gonzo, Riccardo and Shi, Canxin",
    title = "{Dirac brackets for classical radiative observables}",
    eprint = "2506.03249",
    archivePrefix = "arXiv",
    primaryClass = "hep-th",
    reportNumber = "QMUL-PH-25-20",
    doi = "10.1103/ykgq-jqd5",
    journal = "Phys. Rev. D",
    volume = "112",
    number = "10",
    pages = "104060",
    year = "2025"
}

@article{Barcaro:2025ifi,
    author = "Barcaro, Damiano and Del Duca, Vittorio",
    title = "{The central emission vertex of two gravitons}",
    eprint = "2506.11822",
    archivePrefix = "arXiv",
    primaryClass = "hep-th",
    doi = "10.1007/JHEP09(2025)041",
    journal = "JHEP",
    volume = "09",
    pages = "041",
    year = "2025"
}

@article{DelDuca:1995zy,
    author = "Del Duca, Vittorio",
    title = "{Equivalence of the Parke-Taylor and the Fadin-Kuraev-Lipatov amplitudes in the high-energy limit}",
    eprint = "hep-ph/9503340",
    archivePrefix = "arXiv",
    reportNumber = "DESY-95-040",
    doi = "10.1103/PhysRevD.52.1527",
    journal = "Phys. Rev. D",
    volume = "52",
    pages = "1527--1534",
    year = "1995"
}

@article{Lipatov:1991nf,
    author = "Lipatov, L. N.",
    title = "{High-energy scattering in QCD and in quantum gravity and two-dimensional field theories}",
    reportNumber = "IPNO-TH-91-21",
    doi = "10.1016/0550-3213(91)90512-V",
    journal = "Nucl. Phys. B",
    volume = "365",
    pages = "614--632",
    year = "1991"
}

@article{Caola:2021izf,
    author = "Caola, Fabrizio and Chakraborty, Amlan and Gambuti, Giulio and von Manteuffel, Andreas and Tancredi, Lorenzo",
    title = "{Three-Loop Gluon Scattering in QCD and the Gluon Regge Trajectory}",
    eprint = "2112.11097",
    archivePrefix = "arXiv",
    primaryClass = "hep-ph",
    reportNumber = "OUTP-21-28P, MSUHEP-21-035, TUM-HEP-1382/21",
    doi = "10.1103/PhysRevLett.128.212001",
    journal = "Phys. Rev. Lett.",
    volume = "128",
    number = "21",
    pages = "212001",
    year = "2022"
}

@article{DelDuca:2021vjq,
    author = "Del Duca, Vittorio and Marzucca, Robin and Verbeek, Bram",
    title = "{The gluon Regge trajectory at three loops from planar Yang-Mills theory}",
    eprint = "2111.14265",
    archivePrefix = "arXiv",
    primaryClass = "hep-ph",
    reportNumber = "UUIPT-59/21",
    doi = "10.1007/JHEP01(2022)149",
    journal = "JHEP",
    volume = "01",
    pages = "149",
    year = "2022"
}

@article{Falcioni:2021dgr,
    author = "Falcioni, Giulio and Gardi, Einan and Maher, Niamh and Milloy, Calum and Vernazza, Leonardo",
    title = "{Disentangling the Regge Cut and Regge Pole in Perturbative QCD}",
    eprint = "2112.11098",
    archivePrefix = "arXiv",
    primaryClass = "hep-ph",
    reportNumber = "CERN-TH-2021-225",
    doi = "10.1103/PhysRevLett.128.132001",
    journal = "Phys. Rev. Lett.",
    volume = "128",
    number = "13",
    pages = "132001",
    year = "2022"
}

\end{document}